\documentclass[pdftex,twocolumn,epjc3]{svjour3}          

\RequirePackage[T1]{fontenc}

\smartqed  

\RequirePackage{graphicx}
\RequirePackage{mathptmx}      
\RequirePackage{flushend}
\RequirePackage[numbers,sort&compress]{natbib}
\RequirePackage[colorlinks,citecolor=blue,urlcolor=blue,linkcolor=blue]{hyperref}
\usepackage{amsmath,amssymb}
\usepackage{xspace}
\usepackage{lipsum}
\usepackage{upgreek}
\usepackage{float}
\usepackage{bbold}
\usepackage{placeins}
\usepackage{amssymb}
\usepackage{cuted}
\usepackage{subfigure}
\usepackage{multirow}

\usepackage{lineno}
\usepackage{microtype} 

\journalname{Eur. Phys. J. C}

\begin{document}
\newcommand{\pp}           {pp\xspace}
\newcommand{\ppbar}        {\mbox{$\mathrm {p--\overline{p}}$}\xspace}
\newcommand{\XeXe}         {\mbox{Xe--Xe}\xspace}
\newcommand{\PbPb}         {\mbox{Pb--Pb}\xspace}
\newcommand{\pAa}           {\mbox{pA}\xspace}
\newcommand{\pPb}          {\mbox{p--Pb}\xspace}
\newcommand{\AuAu}         {\mbox{Au--Au}\xspace}
\newcommand{\dAu}          {\mbox{d--Au}\xspace}
\newcommand{\pP}{\ensuremath{\mbox{p--p}}\,}
\newcommand{\kstar}        {\ensuremath{k^{\ast}}\xspace}
\newcommand{\rstar}     {\ensuremath{r^{*}}\xspace}
\newcommand{\qstar}        {\ensuremath{q^*}\xspace}
\newcommand{\snn}          {\ensuremath{\sqrt{s_{\mathrm{NN}}}}\xspace}
\newcommand{\pt}           {\ensuremath{p_{\rm T}}\xspace}
\newcommand{\meanpt}       {$\langle p_{\mathrm{T}}\rangle$\xspace}
\newcommand{\ycms}         {\ensuremath{y_{\rm CMS}}\xspace}
\newcommand{\ylab}         {\ensuremath{y_{\rm lab}}\xspace}
\newcommand{\etarange}[1]  {\mbox{$\left | \eta \right |~<~#1$}}
\newcommand{\yrange}[1]    {\mbox{$\left | y \right |~<~#1$}}
\newcommand{\dndy}         {\ensuremath{\mathrm{d}N_\mathrm{ch}/\mathrm{d}y}\xspace}
\newcommand{\dndeta}       {\ensuremath{\mathrm{d}N_\mathrm{ch}/\mathrm{d}\eta}\xspace}
\newcommand{\avdndeta}     {\ensuremath{\langle\dndeta\rangle}\xspace}
\newcommand{\dNdy}         {\ensuremath{\mathrm{d}N_\mathrm{ch}/\mathrm{d}y}\xspace}
\newcommand{\Npart}        {\ensuremath{N_\mathrm{part}}\xspace}
\newcommand{\Ncoll}        {\ensuremath{N_\mathrm{coll}}\xspace}
\newcommand{\dEdx}         {\ensuremath{\textrm{d}E/\textrm{d}x}\xspace}
\newcommand{\RpPb}         {\ensuremath{R_{\rm pPb}}\xspace}
\newcommand{\nineH}        {$\sqrt{s}~=~0.9$~Te\kern-.1emV\xspace}
\newcommand{\seven}        {$\sqrt{s}~=~7$~Te\kern-.1emV\xspace}
\newcommand{\onethree}        {$\sqrt{s}~=~13$~Te\kern-.1emV\xspace}
\newcommand{\twoH}         {$\sqrt{s}~=~0.2$~Te\kern-.1emV\xspace}
\newcommand{\twosevensix}  {$\sqrt{s}~=~2.76$~Te\kern-.1emV\xspace}
\newcommand{\five}         {$\sqrt{s}~=~5.02$~Te\kern-.1emV\xspace}
\newcommand{\twosevensixnn}{$\sqrt{s_{\mathrm{NN}}}~=~2.76$~Te\kern-.1emV\xspace}
\newcommand{\fivenn}       {$\sqrt{s_{\mathrm{NN}}}~=~5.02$~Te\kern-.1emV\xspace}
\newcommand{\LT}           {L{\'e}vy-Tsallis\xspace}
\newcommand{\GeVc}         {Ge\kern-.1emV/$c$\xspace}
\newcommand{\MeVc}         {Me\kern-.1emV/$c$\xspace}
\newcommand{\GeVmass}      {Ge\kern-.1emV/$c^2$\xspace}
\newcommand{\MeVmass}      {Me\kern-.1emV/$c^2$\xspace}
\newcommand{\lumi}         {\ensuremath{\mathcal{L}}\xspace}
\newcommand{\ITS}          {\rm{ITS}\xspace}
\newcommand{\TOF}          {\rm{TOF}\xspace}
\newcommand{\ZDC}          {\rm{ZDC}\xspace}
\newcommand{\ZDCs}         {\rm{ZDCs}\xspace}
\newcommand{\ZNA}          {\rm{ZNA}\xspace}
\newcommand{\ZNC}          {\rm{ZNC}\xspace}
\newcommand{\SPD}          {\rm{SPD}\xspace}
\newcommand{\SDD}          {\rm{SDD}\xspace}
\newcommand{\SSD}          {\rm{SSD}\xspace}
\newcommand{\TPC}          {\rm{TPC}\xspace}
\newcommand{\TRD}          {\rm{TRD}\xspace}
\newcommand{\VZERO}        {\rm{V0}\xspace}
\newcommand{\VZEROA}       {\rm{V0A}\xspace}
\newcommand{\VZEROC}       {\rm{V0C}\xspace}
\newcommand{\Vdecay} 	   {\ensuremath{V^{0}}\xspace}
\newcommand{\ee}           {\ensuremath{e^{+}e^{-}}} 
\newcommand{\pip}          {\ensuremath{\pi^{+}}\xspace}
\newcommand{\pim}          {\ensuremath{\pi^{-}}\xspace}
\newcommand{\kap}          {\ensuremath{\rm{K}^{+}}\xspace}
\newcommand{\kam}          {\ensuremath{\rm{K}^{-}}\xspace}
\newcommand{\pbar}         {\ensuremath{\rm\overline{p}}\xspace}
\newcommand{\kzeros}        {\ensuremath{{\rm K}^{0}_{\rm{S}}}\xspace}
\newcommand{\kzerobar}     {\ensuremath{\rm \overline{K}^0}}
\newcommand{\lmb}          {\ensuremath{\Lambda}\xspace}
\newcommand{\almb}         {\ensuremath{\overline{\Lambda}}\xspace}
\newcommand{\prot}         {\ensuremath{\rm{p}}\xspace}
\newcommand{\aprot}         {\ensuremath{\rm{\overline{p}}}\xspace}
\newcommand{\n}         {\ensuremath{\rm{n}}\xspace}
\newcommand{\an}         {\ensuremath{\rm{\overline{n}}}\xspace}
\newcommand{\kbar}         {\ensuremath{\rm\overline{K}}\xspace}
\newcommand{\Om}           {\ensuremath{\Omega^-}\xspace}
\newcommand{\Mo}           {\ensuremath{\overline{\Omega}^+}\xspace}
\newcommand{\X}            {\ensuremath{\Xi^-}\xspace}
\newcommand{\Ix}           {\ensuremath{\overline{\Xi}^+}\xspace}
\newcommand{\Xis}          {\ensuremath{\Xi^{\pm}}\xspace}
\newcommand{\Oms}          {\ensuremath{\Omega^{\pm}}\xspace}
\newcommand{\SigZ}            {\ensuremath{\Sigma^0}\xspace}
\newcommand{\aSigZ}            {\ensuremath{\overline{\Sigma^0}}\xspace}
\newcommand{\antik}   {$\mathrm{\overline{K}}\,$} 
\newcommand{\km}           {\ensuremath{\rm{K}^{-}}}
\newcommand{\Ledn}         {Lednick\'y--Lyuboshits\xspace}
\newcommand{\SE}         {Schr\"odinger equation\xspace}

\newcommand{\chiEFT}       {\ensuremath{\chi}\rm{EFT}\xspace}
\newcommand{\ks}     {\ensuremath{k^{*}}\xspace}
\newcommand{\rs}     {\ensuremath{r^{*}}\xspace}
\newcommand{\mt}     {\ensuremath{m_{\mathrm{T}}}\xspace}
\newcommand{\Cth}           {C_\mathrm{th}\xspace}
\newcommand{\Cexp}           {C_\mathrm{exp}\xspace}
\newcommand{\CF}           {\ensuremath{C(\ks)}\xspace}
\newcommand{\Sr}            {\ensuremath{S(\rs)}\xspace}
\newcommand{\BBar}            {\ensuremath{\rm{B}\mbox{--}\rm{\overline{B}}}\xspace}
\newcommand{\SPi}         {\ensuremath{\uppi\Sigma}\xspace}
\newcommand{\kbarN}         {\ensuremath{\rm\overline{K}N}\xspace}
\newcommand{\kMinProt}         {\ensuremath{\rm K^{-}p}\xspace}
\newcommand{\LL}            {\ensuremath{\lmb\mbox{--}\lmb}\xspace}
\newcommand{\pprot}            {\ensuremath{\prot\mbox{--}\prot}\xspace}
\newcommand{\LK}{${\mathrm{K}} \Lambda$\xspace} 
\newcommand{\LAK}{${\mathrm{\overline{K}}} \Lambda$\xspace} 
\newcommand{\XiPi}{${\uppi} \Xi$\xspace} 
\newcommand{\SIK}{${\mathrm{\overline{K}}} \Sigma$\xspace} 
\newcommand{\XiEta}{${\eta} \Xi$\xspace} 
\newcommand{\LKzero}{$\mbox{--}{\mathrm{K^0 _S}} \Lambda$\xspace} 
\newcommand{\LKMin}{\ensuremath{\rm K^{-}\mbox{--}\Lambda}\xspace}
\newcommand{\LAL}            {\ensuremath{\lmb\mbox{--}\overline{\lmb}}\xspace}
\newcommand{\pAL}            {\ensuremath{p\mbox{--}\overline{\lmb}}\xspace}

\newcommand{\pAp}            {\ensuremath{p\mbox{--}\bar{p}}\xspace}

\newcommand{\DK}{\ensuremath{\rm D^{+}\mbox{--}K^{+}}\xspace}

\newcommand{\PiMinXiZ}{\ensuremath{\rm \uppi^- \Xi^0 }\xspace}
\newcommand{\PiZXiMin}{\ensuremath{\rm \uppi^0 \Xi^- }\xspace}
\newcommand{\KMinSigZ}{\ensuremath{\rm K^- \Sigma^0}\xspace}
\newcommand{\KZbarSigMin}{\ensuremath{ \overline{\rm K}^0 \Sigma^-}\xspace}
\newcommand{\EtaXiMin}{\ensuremath{\rm \eta \Xi^- }\xspace}
\newcommand{\Imscatt}            {\ensuremath{\Im a_0}\xspace}
\newcommand{\Rescatt}            {\ensuremath{\Re a_0}\xspace}
\newcommand{\effran}            {\ensuremath{d_0}\xspace}
\newcommand{\ImscattAmpl}            {\ensuremath{\mathcal{I}f}\xspace}
\newcommand{\RescattAmpl}            {\ensuremath{\mathcal{R}f}\xspace}
\newcommand{\XRes}          {\ensuremath{\Xi\mathrm{(1620)}}\xspace}
\newcommand{\XResNovanta}          {\ensuremath{\Xi\mathrm{(1690)}}\xspace}
\newcommand{\XResVenti}          {\ensuremath{\Xi\mathrm{(1820)}}\xspace}
\newcommand{\temp}  {\ensuremath{T_{\rm ch}}\xspace}

\newcommand{\Vopt}  {\ensuremath{V_{\rm opt}(r)}\xspace}
\title{Novel constraints on \LAL and \pAL interactions using correlation data}
\subtitle{iCATS: a framework to study correlation functions using optical potentials}

\author{Valentina Mantovani Sarti\thanks{e-mail: valentina.mantovani-sarti@tum.de}}
\institute{Physics Department, Technische Universit\"at M\"unchen, James-Franck-Str., 85748 Garching, Germany\label{addr1}}

\date{Received: date / Accepted: date}

\maketitle

\begin{abstract}
The interaction between baryons and antibaryons (\BBar) remains a fundamental topic in hadronic physics, particularly due to its potential to reveal exotic bound states such as baryonia. While the proton-antiproton (\pAp) system has been extensively studied—mainly via scattering experiments—the interactions involving antihyperons, such as proton-antilambda (\pAL) and lambda-antilambda (\LAL), are still poorly constrained due to the scarcity of experimental data. High-precision data on these systems, covering the low-energy region down to zero momentum, have recently been obtained via femtoscopic measurements in small colliding systems at the LHC. These results offer a unique opportunity to probe these interactions at short distances with unprecedented precision.\\
In this work, we extract for the first time the scattering parameters for the \LAL and \pAL systems by exploiting high-precision correlation data measured by the ALICE experiment in \pp\ collisions. The goal is to investigate potential differences in the interplay between the elastic and annihilation components of the interaction potential. We employ a novel correlation analysis framework, the \textit{imaginary CATS} (iCATS) package, which solves exactly the Schr\"odinger equation with complex optical potentials to model the strong final-state interactions (FSI) in systems dominated by inelastic processes, as is typical in the \BBar case.\\
Our results indicate distinct annihilation characteristics between the \pAL and \LAL systems, challenging the assumption of a universal \BBar dynamics. The extracted scattering amplitudes are compared to available production cross sections and invariant mass spectra involving \pAL and \LAL pairs. The \LAL and \pAL FSI constrained to femtoscopic data deliver an overall consistent agreement for all these observables, indicating the possibility to explore the \BBar sector more in details in future correlation measurements.

\end{abstract}

\section{Introduction}\label{sec:introduction}
Historically, the \BBar interaction has always been a candidate for searches of exotic bound states, so-called baryonia~\cite{Klempt:2002ap,KlemptAnnihilation}. So far, roughly 50 years from the first measurements on the \pAp interaction~\cite{Daresbury-Mainz-TRIUMF:PRELEAR1,Izycki:PRELEAR2,Lindenmuth:PRELEAR3,LEAR1,LEAR2}, there are no conclusive evidences of such states. The challenge lies in the assessment of the annihilation component of the interaction and on its relative strength with respect to the elastic one, fundamental to establish a possible bound spectrum. 
A wide experimental effort, reaching high-precision with the Low Energy Antiproton Ring (LEAR) era at CERN~\cite{LEAR1,LEAR2}, went into the \pAp system. The experiments at the LEAR facility delivered a wealth of scattering data on total, charge-exchange and annihilation cross-sections for antiprotons laboratory momenta starting from $\sim 200$ MeV~\cite{Scatt1,Scatt2,papScatt1,papScattObelixLowMomenta}.The measurements of the shifts in the energy levels and widths of the antiproton hydrogen at thresholdß provided constraints on the spin-averaged scattering parameters, confirming a non-zero imaginary part associated to the annihilation processes~\cite{Klempt:2002ap}. 
These large amount of experimental constraints at low-energy revealed immediately the dominant feature of the \pAp interaction: the annihilation into multi-mesons. The latter, involving multiparticle processes and expected to be a general feature of all \BBar dynamics, is extremely difficult to model. However,  optical potentials have been typically employed in several theoretical approaches, and with different level of complexity, in order to gain a deeper and more quantitative understanding of the annihilation strength and range (see Ref~\cite{Klempt:2002ap,KlemptAnnihilation} for a review). The availability of a large and differential \pAp dataset delivered the possibility to perform energy-dependent partial wave analysis~\cite{Zhou:PWApantip}, although still using optical potentials to phenomenologically model the short-range absorption part.\\
When moving to \BBar data involving antihyperons ($\bar{Y}$), the experimental input is heavily reduced or not available at all. The interaction between a \lmb hyperon and its antiparticle $\bar{\lmb}$ has been prevalently constrained by strangeness-exchange $\pAp \rightarrow \LAL$ cross-section data measured by the PS185 collaboration at LEAR~\cite{Klempt:2002ap}. No scattering data are present so far for the interaction of a proton with a $\bar{\lmb}$ baryon.\\
The LEAR facility was closed in 1996, and since then no dedicated scattering experiments aimed at exploring the \BBar dynamics have been carried out. Nevertheless, the interest on the underlying physics revolving around the \BBar interaction with strange particles has been kept high thanks to measurements of \LAL invariant mass spectra from three-body decays of $B$~mesons~\cite{Belle:LAL1, BaBar:LALB} and charmonia~\cite{BESIII:JPsiLAL,BESIII:3868LAL1,BESIII:3868LAL2} into a neutral meson and \LAL pair. Data on production cross sections $e^+ e^- \rightarrow \phi \LAL, \eta \LAL$ measured at electron-positron colliders are available~\cite{BESIII:LALeta,BESIII:LALPhi}. Important to mention in the context of bound states searches is the observation of threshold enhancements in the reaction $e^+ e^- \rightarrow \LAL$ by BESIII~\cite{BESIII:2018,BESIII:2023}, which triggered many theoretical studies introducing unobserved narrow resonances at threshold and possible sub-threshold \LAL states to explain these data~\cite{Li:Narrow,Salnikov:LALBS}.\\
The \pAL system can also benefit from similar measurements~\cite{BES:JPsipAL,BESIII:pALChi,BESIII:3686pAL}, however, with a much reduced statistics. The lack of data on this \BBar pair translates into difficulties on which assumptions should be made for this system. Several works used, for example, the working hypothesis that the \pAL final-state-interaction (FSI) is the same as the one of the more constrained \LAL system~\cite{Haidenbauer:pAL}. The current large uncertainties on the above-mentioned \pAL spectra prevent any quantitative conclusion on the goodness of such an assumption. More precise data, however still preliminary, on \pAL, \LAL and \pAp differential cross-sections measured in photoproduction experiments by GlueX~\cite{Pauli:GlueX,Li:Gluex} have been presented. This input can help in providing more insights into the possible differences amongst these pairs, however the modeling of such data requires the knowledge of the different production mechanism, currently poorly known.\\
Femtoscopic measurements performed at ultra-relativistic energies have the possibility to overcome the above-mentioned experimental limitations in terms of \BBar pair yields and of providing a less model-independent access to the FSI~\cite{femtoreview,Lisa:2005dd}. Indeed, correlation data obtained in small colliding systems at LHC, such as \pp collisions, delivered already many high-precision measurements on different pairs and proved to be sensitive to the strong force. The reason lies in the fact that, in \pp collisions, particles are emitted from a source with a size of the order of 1 fm~\cite{ALICE:Source, ALICE:Source_pionpion, ALICE:CommonSourceppion, CECA}. This means that the distance between the two particles in the pair is of maximum a couple of fm, exactly within the typical range of the strong interaction. The access to the pair FSI requires though the knowledge of the size of this emitting source~\cite{Lisa:2005dd}. To constrain the latter in a data-driven way, the ALICE collaboration developed a model in which a gaussian core is assumed, common to all hadron pairs and depending only on the transverse mass \mt of the pair, and contributions coming from strong resonance decays feeding into the particles pair lead to non-gaussian tails. The core source size has been extracted as a function of \mt using directly measured proton-proton pairs~\cite{ALICE:Source,ALICE:Source_Corrigendum}, whose interaction is well known and hence it can be considered a "standard hadronic interaction candle". The same approach has been tested on other "standard candles" such as $\pi^\pm \pi^\pm$~\cite{ALICE:Source_pionpion}, $K^+p$~\cite{ALICE:Source_pionpion} and recently on $p\pi^\pm$ pairs~\cite{ALICE:CommonSourceppion} demonstrating a common scaling of the core size. This allowed to access many interactions: from hyperon-nucleon~\cite{ALICE:pL,ALICE:pXi,ALICE:pOmega}, to hyperon-hyperon pairs~\cite{ALICE:LL,ALICE:LXi}, extending to the meson-baryon and meson-meson sector both with strangeness and charm content~\cite{ALICE:pKpp,ALICE:pKCoupled,ALICE:LAKpp,ALICE:DKpp,ALICE:pphi,ALICE:pD}.\\
Measurements by ALICE also on of \pAp, \pAL and \LAL correlations, performed both in \pp~\cite{ALICE:BBarpp} and \PbPb ~\cite{ALICE:BBarPbPb} collisions, delivered high-precision data down to zero momenta. The scattering parameters, by assuming the asymptotic behaviour of the pair wave function via the \Ledn approach~\cite{Lednicky} in all three systems, were extracted from the \PbPb correlation data. The results in \pp collisions, with a larger sensitivity to the short-range dynamics, confirmed the role of multi-meson annihilation in the \pAp system and highlighted a significant tension in reproducing the \pAL correlation using the \Ledn approach with the FSI parameters obtained in \PbPb. This incompatibility raised the question, already addressed in other phenomenological studies~\cite{Kisiel:STARReanalysis}, whether the annihilation dynamics and in general the strong interaction should be similar in all \BBar sectors and whether its properties depend on the strangeness content of the system.\\
In this work we focus on the \LAL and \pAL correlation data measured in the \pp datasample by ALICE~\cite{ALICE:BBarpp}. Our goal is to investigate the FSI of these two pairs by introducing a novel correlation analysis tool, iCATS, able to solve exactly the \SE with complex potentials and to evaluate the corresponding theoretical correlation. The iCATS package represents an extension to the "Correlation Analysis Tool using the \SE equation" (CATS)~\cite{CATS}. By employing optical potentials, we will perform the fits on the available \LAL and \pAL correlations to determine the potential parameters and infer on a possible different annihilation dynamics between the two pairs. For the first time, we deliver the scattering parameters for the \LAL and \pAL interaction using optical potentials, allowing to overcome the limitations of the \Ledn approach used so far in small systems for these pairs. Comparison of the extracted FSI amplitudes to available production cross-section data and invariant mass spectra will be done, showing an overall compatibility between correlation constraints and other experimental observables.

The paper is organized as follows. In Sec.~\ref{sec:formalism} we first present the femtoscopy formalism. An overview with details on the iCATS algorithm to solve the \SE equation in the presence of a complex potential is given in Sec.~\ref{subsec:iCATS}. In Sec.~\ref{subsec:modeling} we describe the modeling of FSI for both pairs and the fitting procedure.
Finally in Sec.~\ref{sec:Results} we show the results on the fit to the measured \LAL and \pAL data, with additional comparison to other experimental observables sensitive to the FSI. Conclusions and future outlooks are given in Sec.~\ref{sec:conclusions}.

\section{Formalism}\label{sec:formalism}
\subsection{Femtoscopy}\label{subsec:femtoscopy}
The main observable in two-body femtoscopy is the correlation function \CF, defined via the Koonin-Pratt equation~\cite{Lisa:2005dd}
\begin{align}\label{eq:KooninPratt}
    C(\ks)=\int S(\mathbf{r}) |\Psi_{\ks} (\mathbf{r})|^2 d^3r.
\end{align}
Here $\ks$ and $r$ are the relative distance and momentum  of the pair, evaluated in the pair rest frame. The emitting source $S(\mathbf{r})$ represents the distribution in $r$ of the two particles in the pair and its size depends on the colliding system in which the measurement is performed~\cite{femtoreview}, as discussed in the introduction. The interaction experienced between the particles in the pair is embedded in the relative wave function $\Psi_{\ks} (\mathbf{r})$. The latter can be obtained exactly by solving the \SE for different values of \ks or by assuming an approximate expression given by its asymptotic behavior at large $r$ within the \Ledn approach. In this last case, by performing the integral in Eq.~\ref{eq:KooninPratt} with a Gaussian source, the \Ledn analytical formula is obtained~\cite{Lednicky}, widely used in heavy-ion collisions where source sizes of above 5 fm are achieved. The \Ledn model suffers of several shortcomings when applied to small source sizes, as the ones obtained in \pp collisions. First, the model is limited to only s-waves. Secondly, the applied correction to the original \Ledn formula for small sources depends on the interplay between the effective range parameter $d_0$ and the source size, which might lead to unphysical negative correlations. Recent theoretical works are aiming at improving these limitations~\cite{AlbaladejoFemto,Murase:pWaveLedn,Albaladejo:2025kuv}.\\
A way to overcome the shortcomings presented within the \Ledn approach is to employ the "Correlation Analysis Tool using the \SE equation" (CATS)~\cite{CATS}, able to compute the two-particle correlation function \CF for any central real potential and emission source function. The CATS framework have been used in the majority of recent correlations measured by the ALICE collaboration in \pp collisions~\cite{ALICE:Source,ALICE:BBarpp,ALICE:pOmega,ALICE:pXi,ALICE:LL,ALICE:pL,ALICE:LAKpp,ALICE:pKpp} and also in several phenomenological studies~\cite{Chizzali:2022pjd,CECA,Sarti:LKMin,Mihaylov:pLHaide}.\\
In this work we present for the first time the iCATS package, which introduces in CATS the possibility to solve the \SE with any complex potential and evaluate the two-particle correlation function for any emitting source profile.
\subsection{iCATS overview}\label{subsec:iCATS}
The core of the iCATS package is designed to solve the radial \SE equation in the continuum for energies $E=\frac{\hslash^2 (\ks)^2}{2\mu}$
\begin{align}\label{eq:radSE}
    \frac{d^2u_{\ks,l}}{dr^2}=\left[2\mu(V_l(r)+V_C(r))+\frac{l(l+1)}{r^2}-(\ks) ^2\right]u_{\ks,l}.
\end{align}
Here $u_{\ks,l} = r R_{\ks,l} (r)$ is related to the radial component $R_{\ks,l}$ of the total wave function $\Psi_{\ks} (r,\theta,\phi) = R_{\ks} (r) Y_l ^m (\theta,\phi)$,written in a separable form thanks to the central potential assumption.
The quantity $\mu$ is the reduced mass of the pair, $V_C(r) = \alpha_s \frac{q_1\cdot q_2}{r}$ is the Coulomb potential present for charged systems and $V_l(r)$ represents the strong complex potential. More details on the iCATS solver can be found in~\ref{Appendix A.}.\\
In Eq.~\ref{eq:radSE} we explicitly show the dependence on the pair relative momentum \ks and on the partial wave $l$. For simplicity, in the following discussion, we will omit these indices and refer simply to $u(r)$.\\
The iCATS solver assumes the following ansatz for the radial solution
\begin{align}\label{eq:wfansatz}
    u(r)=u_R(r) + i u_I(r)
\end{align}
which separates the real and imaginary part of the wave function. By substituting this expression in Eq.~\ref{eq:radSE}, we obtain a system of two coupled equations for $u_R(r)$ and $u_I(r)$\footnote{For simplicity, let us consider a neutral pair with no Coulomb involved.}
\begin{align}
    \begin{cases}\label{eq:System}
        u^{''} _R = \left [ 2m V_R +\frac{l(l+1)}{r^2} -k^2\right]u_R +2m V_I u_I \\
        u^{''} _I = \left [ 2m V_R +\frac{l(l+1)}{r^2} -k^2\right]u_I -2m V_I u_R. \\
    \end{cases}
\end{align}
Following the same approach used in CATS, based on the Euler discretization of the second derivative of the wave functions (see Appendix A.2 in~\cite{CATS}), the solution to the system in Eq.~\ref{eq:System} is achieved by applying a numerical \textit{finite difference method} (FDM) in the $r$ space (see Appendix for more details). The boundary condition at $r=0$, which determines the starting value for $u_R(0)$ and $u_I(0)$, is chosen to be equal to the regular Bessel wave function $j_0(\ks r)$ at the origin. The iterative FDM approach is hence applied up to large values of $r$, corresponding to the asymptotic region where the effect of the strong potential is zero.\\
Once convergence is achieved, the numerical wave function  $u(r)$ must be matched to the asymptotic boundary condition at $r\rightarrow \infty$ in order to determine the correct normalization. Depending on the pair at hand, the asymptotic condition is represented either by free spherical waves or by Coulomb waves. In general, the asymptotic form of the numerical solution reads
    \begin{align}\label{eq:asymptoticGeneral}
    u_{\rm asy} \rightarrow Au^+ - B u^- \Rightarrow \dfrac{u_{\rm asy}}{A} \rightarrow (u^+ - \dfrac{B}{A} u^-) = (u^+ - S^\dagger u^-)
\end{align}
with the outgoing ($u^+$) and incoming ($u^-$) asymptotic waves that, in case of neutral/charged pairs read (assuming $l=0$)
\begin{align}\label{eq:asymptoticSpecific}
    u^{\pm} & = \dfrac{e^{\pm i\ks r}}{2ik} = \dfrac{1}{2} i\cdot r\cdot n_0(\ks r)\pm  \dfrac{1}{2}\cdot r \cdot  j_0(\ks r),\,\,\,\text{(w/o Coulomb)}\nonumber \\
    u^{\pm} & = \dfrac{1}{2ik} [G(\ks r) \pm i\cdot F(\ks r)],\,\,\,\text{(Coulomb)}
\end{align}
The function $n_0(\ks r)$ ($j_0(\ks r)$) is the irregular (regular) spherical Bessel solution, while $G(\ks r)$ (F(\ks r)) is the irregular (regular) Coulomb wave functions~\cite{abramowitz}.\\
The complex numbers $A$ and $B$ in Eq.~\ref{eq:asymptoticGeneral} are determined by imposing the matching between the numerical solution of~\ref{eq:radSE} and this asymptotic condition in which the outgoing wave is normalized to unity. In particular, the outgoing condition written for the femtoscopic studies in Eq.~\ref{eq:asymptoticGeneral} provides access to the conjugate of the scattering matrix $S(\ks)$ as a function of \ks. Once $S(\ks)$ is known, both the complex pahse shifts $\delta_l$ and the scattering amplitude $f(\ks) = \frac{S(\ks) -1}{2i\ks}$ can be evaluated within iCATS. \\
Once the iCATS solver provides the numerical solution $u(r)$ for different values of \ks, the CATS basic functionalities are used to provide as output the total correlation function
\begin{align}
C(\ks) = \sum_{\text{states}(I,S,L,J)} w_{(I,S,L,J)}C_{I,S,L,J}(\ks),
\end{align}
where $(I,S,L,J)$ represent the isospin, spin, angular momentum and total angular momentum quantum numbers characterizing the pair under study. Each of the correlations $C_{I,S,L,J}(\ks)$ are evaluated using the Koonin-Pratt formula in Eq.~\ref{eq:KooninPratt} and the corresponding wave function, with a final weight $w_{(I,S,L,J)}$ that accounts for the degeneracy of the states (see Appendix A.2 in~\cite{CATS}).\\
The detailed description on how the iCATS solver is used to extract the scattering parameter for the \pAL and \LAL systems from the measured correlations is described in the next section.

\subsection{Modeling the \pAL and \LAL correlation}\label{subsec:modeling}
\begin{figure}[h!]
\includegraphics[width=\columnwidth]{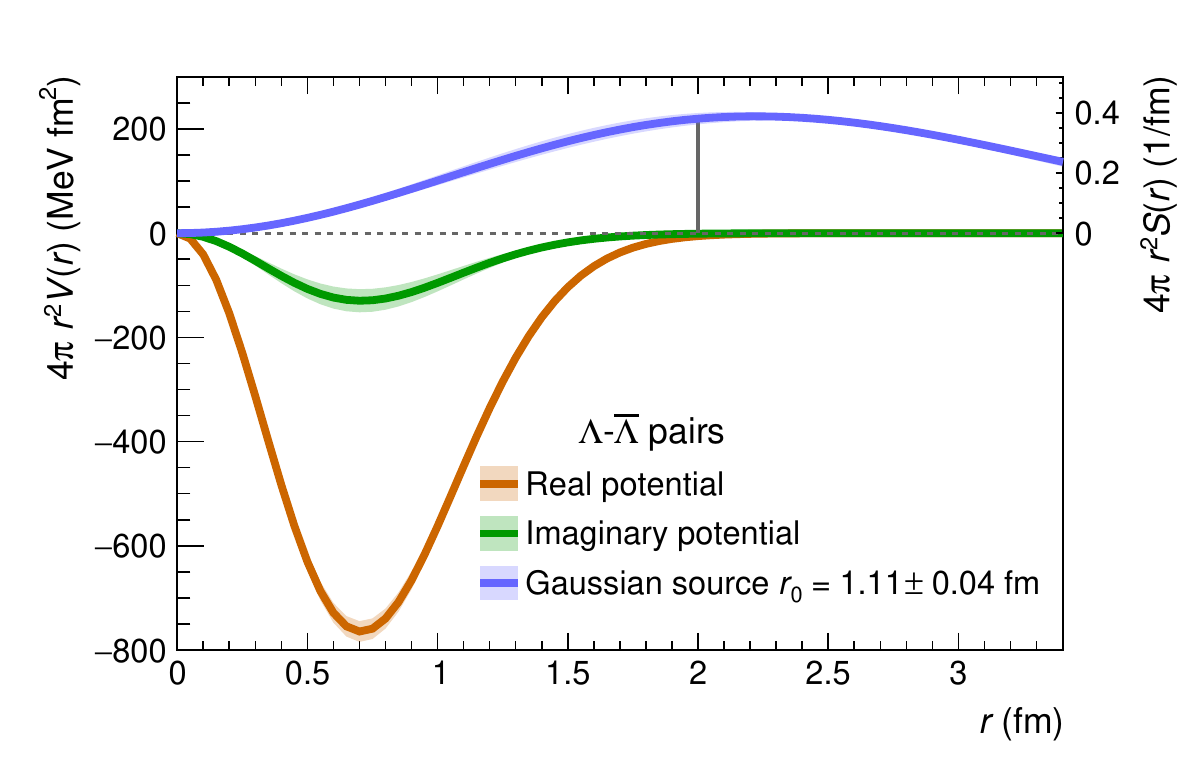}
\caption{(Color online). Real (orange) and imaginary (green) component of the \Vopt potential for the \LAL system obtained in this work. The potential is multiplied by the factor $4\pi r^2$ in order to show the overalpping sensitivity with the emitting source density distribution $4\pi r^2 \cdot S(r)$, shown in blue, for the value of the gaussian source size given for \LAL pairs in~\cite{ALICE:BBarpp}.}
\label{fig:SourcevsPot}
\end{figure}

The measured correlations in \pp collisions of the \LAL and \pAL pairs delivered in~\cite{ALICE:BBarpp} have been modeled using the \Ledn formula, assuming the scattering parameters obtained from the \PbPb femtoscopic data with source sizes of 4 fm and above~\cite{ALICE:BBarPbPb}. The large interparticle distance between the baryon and its antibaryon partner obtained in such large emitting sources should lead to a reduced sensitivity to the short-range annihilation part, typically acting below 1 fm~\cite{Klempt:2002ap,KlemptAnnihilation,Zhou:PWApantip}. The data collected in the high-multiplicity \pp collisions data sample contains instead \LAL and \pAL pairs that are produced also at such short distances, as shown in Fig.~\ref{fig:SourcevsPot} for the \LAL case specifically. In Fig.~\ref{fig:SourcevsPot}, the source density distribution obtained for \LAL pairs, using the source size reported in~\cite{ALICE:BBarpp}, is shown in blue with the band representing the associated uncertainty. The factor $4\pi r^2$ is also used to multiply the profiles of $V_R(r)$ and $V_I(r)$ determined in this work. The \lmb and $\bar{\lmb}$ particles in the pair produced at interparticle distances of $\approx 2$ fm are experiencing the action of the \Vopt potential, and thanks to the source  size achieved in the \pp collision measurements, there is a significant probability to achieve these conditions (see vertical gray line).
In this work, we aim at testing the \LAL and \pAL interactions on correlation data by assuming an optical potential \Vopt with an imaginary component responsible for the annihilation processes. This modeling assumption is one of the most used in many theoretical investigations on \BBar interactions~\cite{BryanPhillips,DoverRichard,Kohno:AnnIntoTwoMesons,Kohno:pAptoLAL,Zhou:PWApantip,HaidenbauerHolindeSpethYantiY}.
The Yukawa-type optical potential reads
\begin{align}\label{eq:YukawaPot}
    \Vopt = \left [ V_{R}+i\cdot V_{I} \right ] \cdot e^{-m_{\rm ex.} ^2 r^2}
\end{align}
where the range of the potential is determined by the inverse of the exchanged meson mass, $m_{\rm ex.}$. Similar phenomenological interaction forms, either based on real-only potentials or complex ones, have been already used in recent femtoscopic studies on p-$\phi$~\cite{Chizzali:pPhiBS}, p-\lmb~\cite{Mihaylov:pLHaide} and D-light mesons~\cite{ALICE:DKpp} pairs. 
We assume the exchange of the lightest scalar meson whose mass is given by $2m_{\pi}$. The considered potential is assumed in s-wave only and for this work, in order to test the robustness of the iCATS framework and to reduce the number of fit parameters, we assume a spin-averaged interaction\footnote{Both \LAL and \pAL pairs have a $^1 S_0$ singlet and a $^3 S_1$ triplet state. If a spin-dependence is included, due to the lack of theoretical and experimental guidelines on possible relations between the interaction of the two pairs, the total number of parameters to be fitted will be  4 for each pair (2 for $V_R$ and 2 for $V_I$), leading to a total of 8 parameters.}. Possible extensions to higher-partial waves and to additional contributions to the assumed central potential (e.g one-pion exchange large $r$ tails) can be done in the future but currently there is no strong theoretical guidance on how to parametrize these terms.\\
The potential strengths for the real $V_R$ and imaginary $V_I$ component are the free fit parameters, in the comparison of the modeled correlation to data.\\
A necessary ingredient in evaluating the theoretical correlation function for the \pAL and \LAL pairs is the source size. As already discussed in detail in~\cite{CATS}, the CATS framework handles both analytical source function, typically based on Gaussian distributions, and also numerically evaluated distributions computed for example from event generators~\cite{EPOS:2013ria, PythiaGuide}. The iCATS package inherit all the CATS functionalities on the emitting source modeling and, following the same approach used in the experimental analysis in~\cite{ALICE:BBarpp}, we adopted a Gaussian source profile for both \pAL and \LAL pairs with the source sizes reported in~\cite{ALICE:BBarpp}: $r_{0,\pAL} = 1.15 \pm 0.04$ fm and $r_{0,\LAL} = 1.11 \pm 0.04$.\\
To fit the measured \pAL and \LAL correlation data, we adopt the same approach of~\cite{ALICE:BBarpp} by using the data and background given in the corresponding public~\href{https://www.hepdata.net/record/ins1862795}{HEP-Data repository}\footnote{https://www.hepdata.net/record/ins1862795}.\\
The total correlation function to be fitted to the \LAL and \pAL correlation data hence reads
\begin{align}\label{eq:totcorrelation}
   & C_\mathrm{tot}(\ks) =  N_D \times C_\mathrm{model}(\ks) \times C_\mathrm{background} (\ks),\nonumber \\
\end{align}
in which $C_\mathrm{background} (\ks)$ represents the background. The normalization $N_D$ is left free to vary in the fit and the fit range is performed for values of \ks up to 350 \MeVc.
The measured correlation functions also include residual correlations, namely correlations between the two particles in the pair in which one is primarily produced and the other stems from long-lived (lifetime $c\tau \gg$ fm) resonance decays~\cite{ALICE:Run1}. The effect of these correlations is to dilute the genuine signal in the experimental correlation when both particles in the pair are primaries. Accordingly, the modeled $C_\mathrm{model}(\kstar)$ correlation to be accounted for in Eq.~\ref{eq:totcorrelation} reads
\begin{align}\label{eq:decompCF}
    & C_{\rm{model}}(\ks) =  1 + \sum_{i} \lambda_{i} \times (C_{i}(\ks) - 1).
\end{align}
The lambda parameters $\lambda_{i}$ are typically reported in the experimental analyses and obtained in a data-driven way by comparing  raw data to Monte Carlo simulated data. More details on the determination of the $\lambda_{i}$ weights can be found in~\cite{ALICE:Run1}. Here, we assume that the genuine \pAL (\LAL) correlations corresponds to $i=0$ and the values are directly taken from~\cite{ALICE:BBarpp}, $\lambda_{\rm gen, \pAL} = 0.458$ ($\lambda_{\rm gen, \LAL} = 0.309$).
The measured \pAL correlation includes also a residual $7\%$ of \LAL correlations ($\lambda_{p_{\lmb}\bar{\lmb}} = 0.07$), when the proton is originating from the decay of a \lmb. This residual correlation has been accounted for in the \pAL fit and modeled assuming the interaction extracted in the iCATS fit to the \LAL pairs. The remaining residuals, accounting mainly for those residuals involving protons and heavier antihyperons ($\bar{\Sigma},\bar{\Xi}$) are considered flat since no theoretical predictions are available and to follow as close as possible the original experimental analysis in~\cite{ALICE:BBarpp}.
In the \LAL case, all the remaining residuals are considered flat.\\
The fitting procedure uses iCATS to model the correlations entering in Eq.~\ref{eq:decompCF} and by fitting simultaneously the \pAL and the \LAL correlations given in~\cite{ALICE:BBarpp} with Eq.~\ref{eq:totcorrelation}, leaving free for each pair the \Vopt parameters and the overall normalization $N_D$. The uncertainties reported on the source radii are included in the fitting approach, along with a variation on the fit range of $\pm10\%$. In order to take into account as well the uncertainties coming from the data and background, we employ a bootstrap sampling~\cite{Bootstrap}, similarly to the experimental analysis~\cite{ALICE:BBarpp}.
\section{Results}\label{sec:Results}
\begin{figure*}[t!]
\begin{center}
\includegraphics[width=\columnwidth]{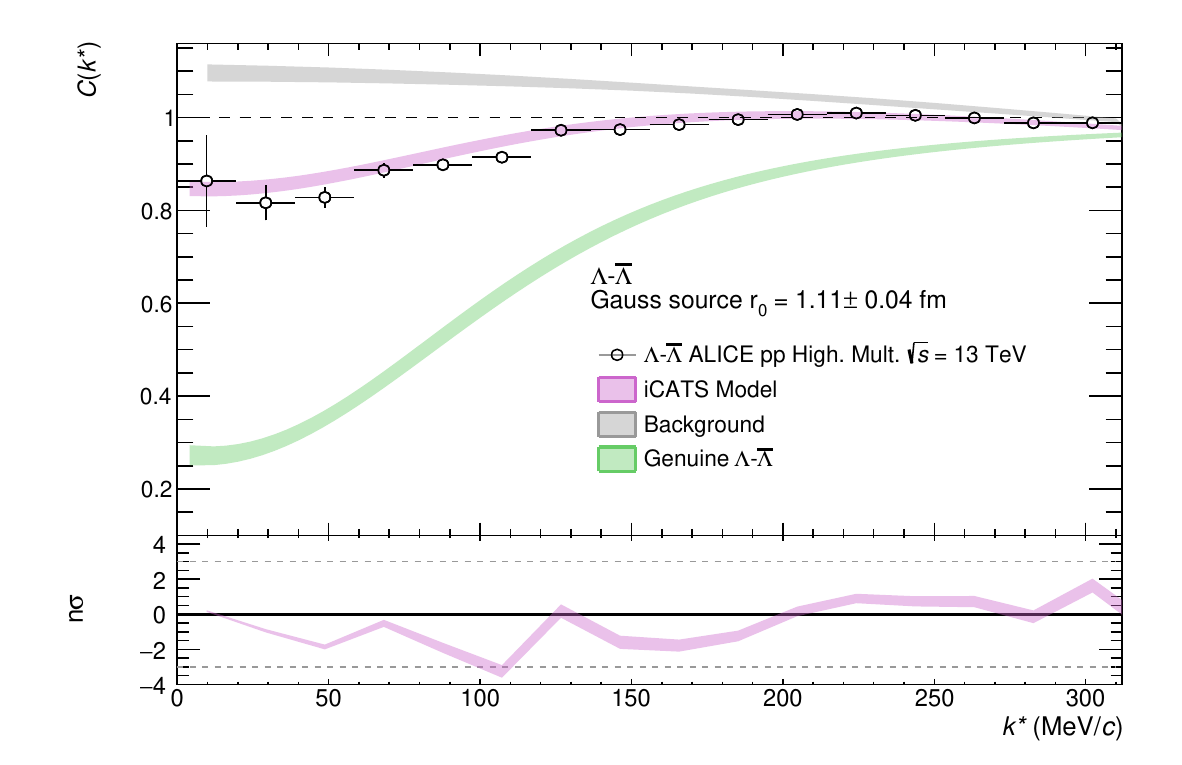}
\includegraphics[width=\columnwidth]{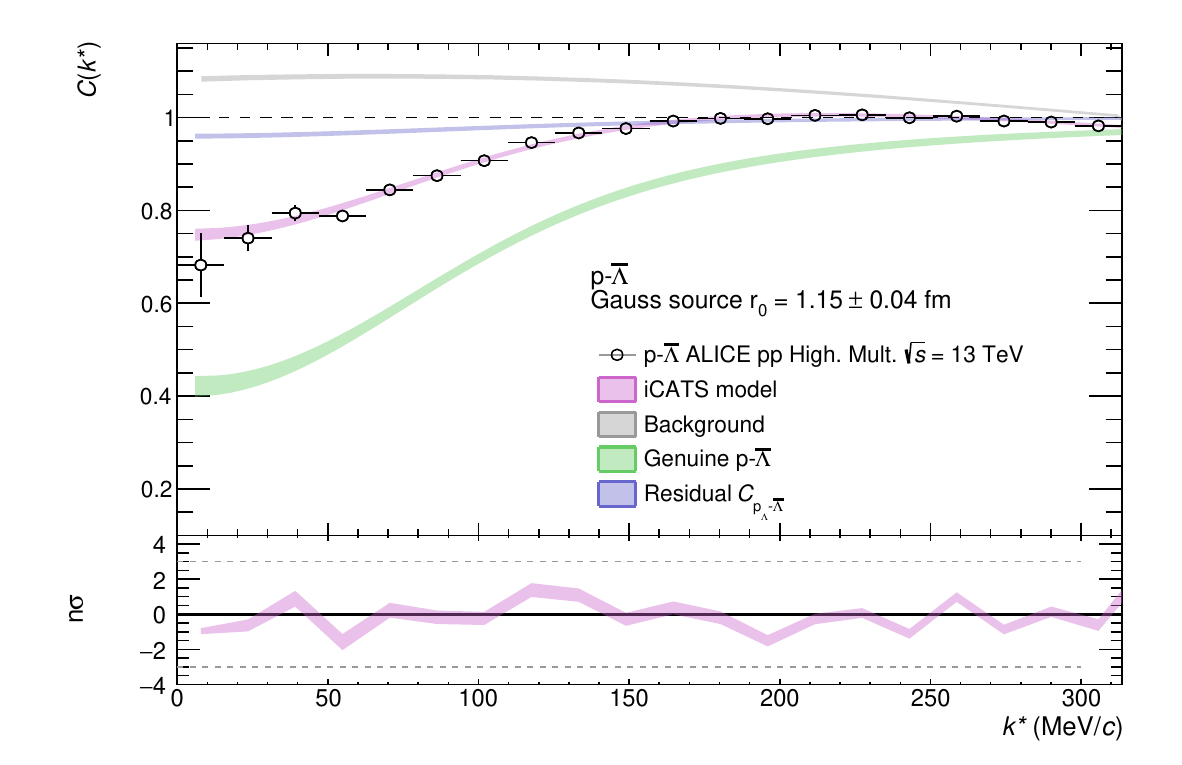}
\caption{(Color online). Results obtained with iCATS for the \LAL (left) and \pAL (right) measured correlation functions. The publicly available data (black empty markers) and background (grey band) correspond to the ones in Fig.3 of Ref.~\cite{ALICE:BBarpp}. The results of the fit using Eq.~\ref{eq:totcorrelation} are shown in pink. The genuine correlations and the modeled residual in the \pAL case are presented in green and blue, respectively. The lower panels deliver the deviation, expressed in terms of standard deviations $\sigma$, of the fit results to the data}
\label{fig:BBarCF_pp}
\end{center}
\end{figure*}

The results of the fitting procedure described in Sec.~\ref{sec:formalism} are presented in Fig.~\ref{fig:BBarCF_pp} for the \LAL and \pAL correlations.

The fits obtained using the iCATS package are able to reproduce the measured correlations in the considered \ks range, as visible from the lower $n\sigma$ panels delivering the deviation between data and model. In the \pAL case, we also report the contribution of the residual \LAL correlation (blue band), which is significantly reduced in strength by both the kinematic transformation to the \pAL frame momentum and by the small $\lambda$ parameter ($\approx 7\%$). The results on the remaining measured correlation in different \mt ranges are shown in Fig.~\ref{figApp:LALCF_pp} and Fig~\ref{figApp:pALCF_pp} in~\ref{Appendix B.}. \\
The genuine \LAL and \pAL correlation function, shown in green, are rather different in strength, reflecting a difference in the underlying FSI since the source sizes of the two systems are compatible. Such a conclusion was qualitatively reached as well in~\cite{ALICE:BBarpp} but no more sophysticated model beyond the \Ledn approach was available to test more in detail such hypothesis.
The iCATS framework, suited to model optical potentials, presented in this work allowed for the first time a more quantitative study of these data.\\

In Tab.~\ref{tab:PotParsScattPars} we report the values of the extracted potential parameters and the corresponding scattering length $a_0$. The modeling via optical potentials with iCATS improved significantly the precision on the scattering parameters extracted for these pairs. The scattering length for the \LAL pairs extracted in \PbPb measurements~\cite{ALICE:BBarPbPb} and used to model the correlations obtained in \pp collisions in~\cite{ALICE:BBarpp} have uncertainties of the order of $\approx20\%$ and $\approx50\%$ for \Rescatt and \Imscatt, respectively.

\begin{table*}[t!]
\centering
\caption{Values of the real $V_R$ and imaginary $V_I$ potentials extracted from the fits to the measured correlations. In the last column, the corresponding scattering lengths are shown.}
\label{tab:PotParsScattPars}
\begin{tabular}{c|c|c|c}
\hline
\hline
Pair     & $V_R$ (MeV) & $V_I$ (MeV) & $a_0$ (fm)  \\ \hline
$\LAL$ & $-331.02\pm 8.63$ & $-56.12\pm9.61$ & $(-1.54\pm0.03)+i\cdot(0.43\pm0.08)$ \\ \hline
$\pAL$ & $-675.35\pm 45.00$ & $-98.92\pm 66.48$ & $(-0.84\pm0.06)+i\cdot(0.16\pm0.09)$ \\ \hline\hline

\end{tabular}
\end{table*}

In the results presented here, a precision of the order of $2-5\%$ is achieved for the real component of the scattering length \Rescatt, while a slightly larger value (max $15\%$) is obtained for the imaginary part \Imscatt. Similar values hold also for the \pAL case, with a slightly higher precision ($\approx 30\%$) achieved in the determination of the \Imscatt in \PbPb. Overall, when compared to the scattering parameters from the \PbPb analysis in~\cite{ALICE:BBarPbPb}, we achieve a compatibility within $3\sigma$ for most of the scattering lengths values. A significant deviation above $3\sigma$ is observed for the \Rescatt of \LAL pairs.
For a more quantitative comparison of the iCATS FSI obtained in this work, we refer the reader to Fig~\ref{figApp:CFPbPb} in~\ref{Appendix B.}, where a nice agreement between the iCATS FSI and the \pAL and \LAL correlation functions measured in \PbPb at \fivenn is achieved.\\
In literature, predictions on the \LAL interaction constrained to the strangeness-exchange cross-section data measured at LEAR~\cite{Klempt:2002ap,BBar3,HaidenbauerHolindeSpethYantiY,Kohno:pAptoLAL,Haidenbauer:pictureLAL} are available. In particular, in Ref.~\cite{Haidenbauer:LAL} where the effect of \LAL FSI on measured near-threshold enhancements is nicely discussed, the scattering lengths\footnote{The sign convention reported in Tab.2 of Ref.~\cite{Haidenbauer:LAL} has the opposite sign with respect to the traditional femtoscopic one employed in this work in which the effective range expansion reads: $k\cdot ctg(\delta_0)=\frac{1}{a_0}+\frac{1}{2}d_0 k^2$} for the \LAL system are reported for different meson-exchange models. The spin-average values of \Rescatt vary approximately from $-0.63, -0.66$ (Model I, Model II) to $-1.1$ fm (Model IV), while the \Imscatt ranges respectively from $0.55$ to $0.83$ fm. The authors in~\cite{Haidenbauer:LAL} deliver in the appendix a qualitative comparison with the \LAL data of Fig.~\ref{fig:BBarCF_pp}. In order to test the sensitivity of \BBar correlation data to the underlying dynamics, we used the iCATS package to generate optical potentials tuned to reproduce the scattering lengths reported in~\cite{Haidenbauer:LAL} and perform a full fit to the data, using the publicly available uncertainties on the source and background. Results are reported in~\ref{Appendix B.}, showing an overall disagreement to the data in the low-intermediate \ks region for Model I and II with a much smaller \Rescatt with respect to the on obtained here. Such a comparison can help in excluding these models from future analyses in the \BBar sector. 

\begin{figure}[b!]
\includegraphics[width=\columnwidth]{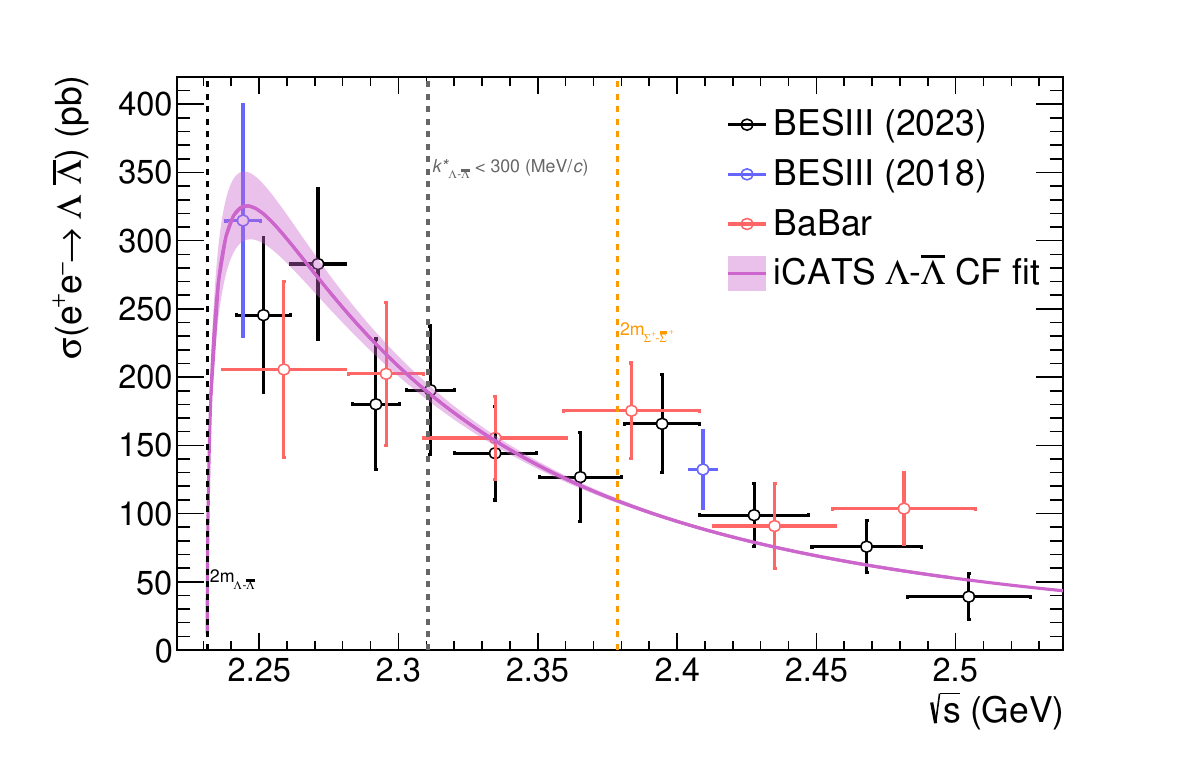}
\caption{(Color online) Cross-section as a function of $\sqrt{\rm s}$ for $e^+e^-\rightarrow \LAL$ measured by BaBar (orange markers)~\cite{BaBar:2007fsu} and BESIII (blue and black markers)~\cite{BESIII:2018,BESIII:2023} compared to the results obtained assuming the \LAL FSI extracted with iCATS from correlation data in~\cite{ALICE:BBarpp}, shown by the pink band. The width of the band corresponds to the 1$\sigma$ total uncertainty on the extracted scattering amplitude. The grey dashed vertical line corresponds to the range in \ks where the \LAL shows a signal of the strong interaction. The orange dashed vertical line indicates the opening of the $\Sigma\bar{\Sigma}$ channel.}
\label{fig:eeLALBesBaBar}
\end{figure}

For the \LAL case, we obtain a situation similar to the one observed in the study of the p-$\phi$ correlation case in~\cite{Chizzali:pPhiBS} in which a negative, large value (in modulus) of \Rescatt was found, compatible with the presence of a bound state in the spin doublet state. In our work, due to the lack of solid constraints on any of the two \LAL spin channels, we obtain spin-averaged scattering parameters which however show a typical indication of a sub-threshold shallow bound state ($E_b \sim 30$ MeV, estimated via the Efimov formula~\cite{Efimov}).
We would like to stress that these are just qualitative indications, since for a full assessment of bound states, a determination of the singlet- and triplet-spin contribution is needed in the future. However, there have been several hints that, particularly in the spin triplet state of the \LAL system, a possible state close to the threshold might be present. The possibility to extract as well the spin-dependence from correlation data could be achieved in more refined studies in the  future by performing a combined fit to production cross-section and invariant mass spectra measured in "spin filtering" reactions (e.g spin triplet in $e^+e^- \rightarrow \LAL$), more sensitive to the singlet or triplet signal (see ~\cite{Haidenbauer:LAL,Haidenbauer:pAL} for more details).\\
An additional crosscheck on the robustness of the results presented here was performed by comparing the FSI extracted from the correlation data with iCATS to available measurements of the \LAL and \pAL cross-sections and invariant mass spectra dominated by the s-wave dynamics. An example of such data is given by the production cross-section $\sigma(e^+ e^- \rightarrow \LAL)$ measured using initial-state radiation. Data on this reaction  collected by BaBar~\cite{BaBar:2007fsu} (orange), BESIII in 2018~\cite{BESIII:2018} (blue) and in 2023~\cite{BESIII:2023} (black) are shown in~Fig.~\ref{fig:eeLALBesBaBar}.

Since the final state \LAL produced in the $e^+e^-$ annihilation is in the triplet state, in order to compare our results to data, we weighted the total scattering amplitude with the corresponding statistical factor. We focus on the role played by the \LAL strong FSI on the $\sigma(e^+ e^- \rightarrow \LAL)$ cross-section. Therefore, the calculated cross-section is proportional to $\frac{4\pi \alpha^2}{3 s}\beta \cdot |A(s)|^2$, where $\alpha$ is the fine-structure constant, $\beta=\sqrt{1-4\frac{m_\Lambda ^2}{s}}$, $s$ is the squared center-of-mass energy and $A(s)$ is the total reaction amplitude~\cite{Haidenbauer:EMTimelike}. Since we aim at testing the iCATS FSI on the energy region close to the \LAL threshold, we estimated the reaction amplitude $A$ via the Migdal-Watson approach~\cite{Watson:1952ji}, assuming that in this energy regime the reaction amplitude is fully dominated by the \LAL scattering amplitude. This approach is rather popular thanks to its simplicity and hence it has been already used in several \BBar theoretical studies~\cite{Sibirtsev:pApJPsi,Kerbikov:2004gs,Bugg:2004rk}. The theoretical cross-sections, caveat a  has been simply rescaled to the data for a straightforward comparison\footnote{}.\\
The resulting cross-section from the iCATS FSI, shown in the pink band in Fig.~\ref{fig:eeLALBesBaBar}, reproduces nicely the region close to the \LAL threshold, up to $\sqrt{\rm s}\approx 2.35$ GeV, corresponding to a \ks value of approximately 370 \MeVc. The enhancement observed at $\sqrt{\rm s}\approx 2.4$ GeV corresponds to the opening of the $\Sigma\bar{\Sigma}$ channel, currently not included in our model and so far in the majority of models for \LAL FSI~\cite{Haidenbauer:EMTimelike}. Theoretical predictions based on meson-exchange models within a full coupled-channel approach~\cite{HaidenbauerHolindeSpethYantiY} show that the transition cross-section \LAL to $\Sigma\bar{\Sigma}$ is largely suppressed with respect to the diagonal one, hence neglecting this coupling should not have a significant effect in the region close to threshold. Moreover, no observation of a cusp structure was reported in the measured \LAL correlations at $\ks\approx 410$, corresponding to the $\Sigma\bar{\Sigma}$ opening, supporting indeed the sub-dominant effect of this transition potential.\\
\begin{figure*}[t!]
\includegraphics[width=\columnwidth]{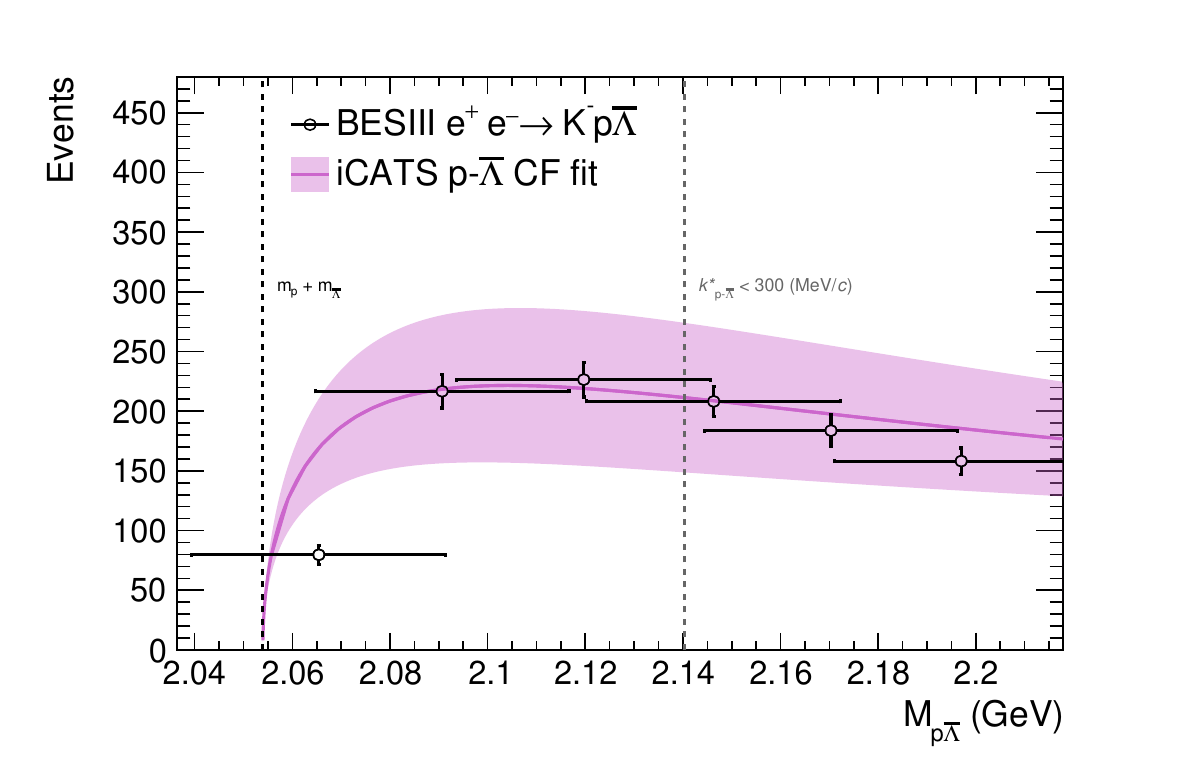}
\includegraphics[width=\columnwidth]{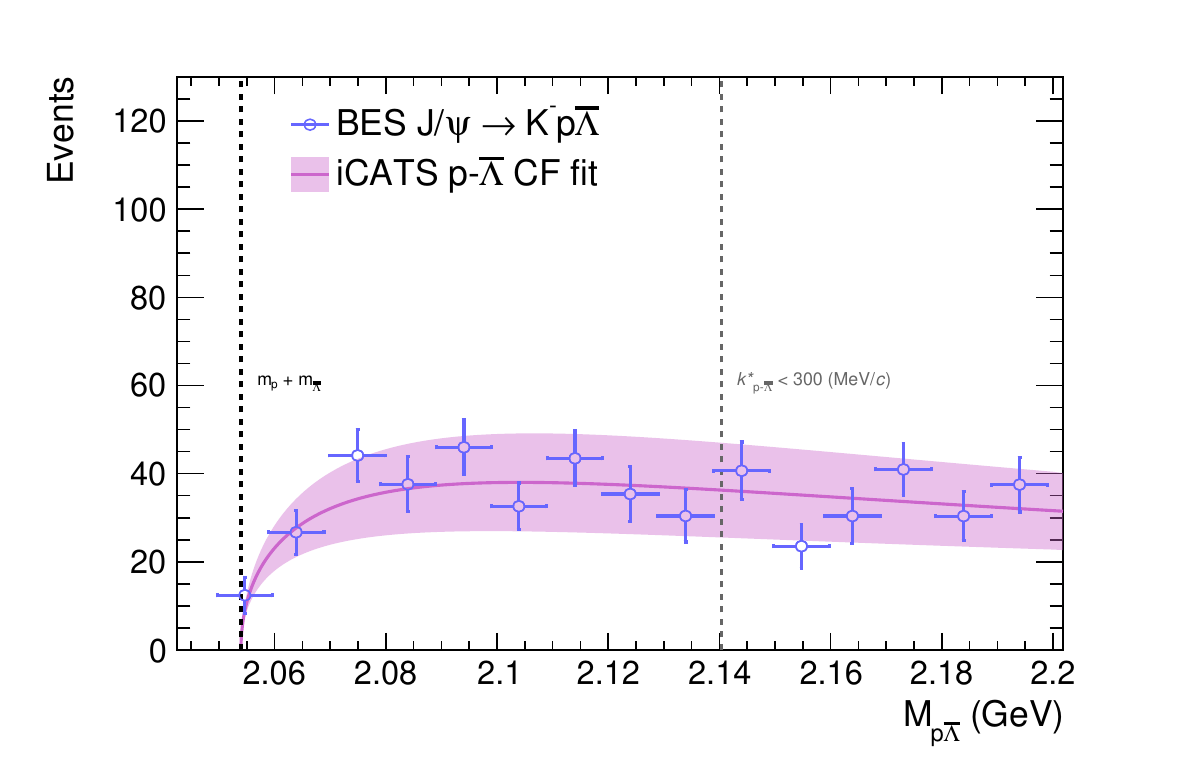}
\caption{(Color online) Invariant \pAL mass spectra from $e^+e^-\rightarrow K^- \pAL$ measured by BESIII (left)~\cite{BESIII:eepAL} and from $J/\Psi\rightarrow K^- p\bar{\lmb}$ BESIII (right)~\cite{BES:JPsipAL} compared to the results obtained assuming the \pAL FSI extracted with iCATS from correlation data in~\cite{ALICE:BBarpp}, shown by the pink band. The grey dashed vertical line corresponds to the range in \ks where the \pAL shows a signal of the strong interaction.}
\label{fig:eepAL}
\end{figure*}
Of particular interest is the \LAL threshold enhancement seen in BESIII, whose trend is confirmed by both the 2018 and 2023 data. Threshold enhancements have been observed in several \pAp invariant mass spectra in the decays of $J/\Psi$~\cite{BES:pApJPsi} and B mesons~\cite{Belle:pApB1,Belle:pApB2}, in which the Coulomb attraction as well has an impact on describing the cross-section increase~\cite{Sibirtsev:pApJPsi}. This increase, observed in the BESIII data close to threshold, is not expected for a neutral pair and it could only arise as a result of the strong \LAL dynamics. It is worth mentioning that several Vector Meson Dominance models managed to reproduce the \LAL threshold enhancement in Fig.~\ref{fig:eeLALBesBaBar} by assuming the presence of unobserved narrow resonances at threshold~\cite{Li:Narrow} or by adding some additional sub-threshold broad vector meson states~\cite{Yang:VDMResoLAL1,Bai:VDMResoLAL2,Xiao:VDMResoLAL3}. A recent work provides also an explanation of the observed \LAL enhancement by linking it to the presence of a state below threshold arising from the underlying \LAL potential~\cite{Salnikov:LALBS}. Interestingly, the authors in~\cite{Salnikov:LALBS} deliver a binding energy for such state of roughly 30 MeV, in line with the qualitative indication we obtain in our analysis.\\
We conclude this first part of the section by highlighting that some of the available meson-exchange models~\cite{Haidenbauer:LAL,Dai:Review2025EMYantiY} anchored to $\pAp \rightarrow \LAL$ data deliver a much flatter behavior of the production cross-section. These models whose scattering lengths are reported in~\cite{Haidenbauer:LAL} and in~\ref{Appendix B.}, show also tension with the description of the measured \LAL correlation functions, see~\ref{Appendix B.}. For completeness, we evaluated the $\pAp \rightarrow \LAL$ cross-section with the iCATS FSI, following the same approach as for the $e^+e^-$ case. Results are shown in~\ref{Appendix B.}. This production cross-section is expected to have, already at threshold, a non-negligible p-wave contribution~\cite{HaidenbauerHolindeSpethYantiY,Kohno:pAptoLAL}, hence we cannot achieve a perfect description between our s-wave modeling and the LEAR data.\\
Following the Migdal-Watson approach used in the \LAL case, we tested the iCATS FSI for the \pAL system on two different sets of data: the \pAL invariant mass spectra obtained from the $J/\Psi\rightarrow K^- p\bar{\lmb}$ decay by BES~\cite{BES:JPsipAL} and the reaction $e^+e^-\rightarrow K^- p\bar{\lmb}$ measured by BESIII~\cite{BESIII:eepAL}. Similarly to the \LAL case, the aim is to focus on the \pAL FSI and its energy dependence close to threshold. Based on this considerations, the estimated the invariant-mass spectra are proportional to $\propto k^\ast _{\pAL} |A(M_{\pAL})|^2$~\cite{Haidenbauer:pAL,Haidenbauer:LAL}, in which the amplitude is provided by iCATS.
Also in the \pAL case, the threshold enhancement observed in both measurements was interpreted with the presence of resonances lying across the \pAL nominal mass with widths of the order of few tens of MeV~\cite{BES:JPsipAL,BESIII:eepAL}. It should be mentioned that correlations proved to be sensitive to the presence of states close to threshold, as shown by the first experimental evidence of the $\Xi(1620)$ in the $\lmb K^-$ channel via the measurements of $\lmb K^-$ correlations in \pp collisions~\cite{ALICE:LAKpp,Sarti:LKMin}. In the \pAL correlation data reported in~\cite{ALICE:BBarpp}, within the current uncertainties, no significant structure can be seen and the data can be nicely reproduced without any additional inclusion of states. Since both these processes are sensitive to the singlet and triplet \pAL states, the total iCATS scattering amplitude was considered in the calculation. The iCATS results are able to overall reproduce the data within their uncertainties, delivering a consistent picture between the extracted \pAL FSI from correlation measurements and these additional observables.\\
We conclude this section by emphasizing that the main goal of this work was to assess the capabilities of a novel correlation analysis package iCATS, particularly suited for hadronic interactions dominated by inelastic channels. The application to the measured \pAL and \LAL was able to successfully demonstrate the improved precision and flexibility of iCATS in modeling the annihilation dynamics via optical potentials.  The results on correlation data shows a different behaviour of the underlying strong dynamics in the two analysed pairs. The FSI extracted from the fit to the \pAL and \LAL correlations provide new constraints on these two poorly known interactions and it shows that, for systems involving antihyperons, femtoscopy can provide complementary support to the renewed interest on \BBar interactions currently being explored in recent photoproduction measurements at JLAB~\cite{Pauli:GlueX} and in future experiments at the CERN Antiproton Decelerator facility~\cite{Caravita:CERNAD}. 

\section{Summary and conclusions}\label{sec:conclusions}
In this work we presented novel constraints on the FSI of \pAL and \LAL systems obtained from measured correlation functions in \pp collisions. The interacting potentials for these two pairs have been modeled with the novel iCATS package, an extension to the CATS framework, widely used so far in correlation data analysis. The iCATS is able to solve the \SE with complex optical potentials and evaluate the theoretical correlation function for any emitting source profile.\\
The iCATS functionalities deliver the exact wavefunction down to zero interparticle distances, hence allowing to access in a more precise manner the annihilation part of the underlying interaction for the \BBar case at hand.\\
The results presented here shows a clear difference in the short-range annihilation dynamics between the \LAL and \pAL systems which might arise from a different strangeness content, leading to different absorptions channels. The FSI interaction for \pAL and \LAL obtained with iCATS was compared to available production cross-sections and invariant mass spectra measured in $e^+e^-$ collisions. An overall compatibility between the extracted scattering amplitude from correlation data and these other observables is achieved, opening the possibility to exploit femtoscopy data to constrain the challenging \BBar interaction. Indication of a possible \LAL bound state, even if based on qualitative estimates on the extracted spin-average scattering parameters, has been observed. However, as already concluded in other investigations, the threshold enhancements in the considered cross-sections and spectra can be reproduced by the iCATS amplitudes without advocating for any additional resonance.\\
Future combined analysis involving \pAL and \LAL correlation measurements and observables dominated by specific spin and partial waves could allow a more precise determination of the spin-dependence of the annihilation dynamics. An improved knowledge on the interplay between the elastic and annihilation part of the \BBar interaction can have significant impact on searches for exotic baryonium states~\cite{Klempt:2002ap,KlemptAnnihilation}, on the determination of electromagnetic hyperon form factors~\cite{Dai:Review2025EMYantiY} and on neutron skin measurements~\cite{Schupp:Skin,PUMA:2022ngr}.\\ Thanks to the possibility of producing in the same amount large yields of multi-strange baryons and antibaryons at LHC energies~\cite{ALICE:EnhancStrange}, along with the high-statistics already achieved in correlation involving $\Xi(\bar{\Xi})$ and $\Omega(\bar{\Omega})$ particles, femtoscopy in the next years will be able to tackle the measurements of many other $Y\bar{Y}$ interactions.

\begin{acknowledgements}
This work was supported by the Deutsche Forschungsgemeinschaft (DFG) through the grant MA $8660/1-1$. The author would like to thank Dr. Mihaylov and Dr. Kamiya for the fruitful discussions. A special thanks as well to Prof. Fabbietti and Dr. Feijoo for the constructive feedback.
\end{acknowledgements}

\vspace{-0.45cm}

\appendix
\section{iCATS solver and computation of the wave function}\label{Appendix A.}


The iCATS package is implemented in the C++ core class of the CATS framework. In this section we would like to address some technical details on the software which can be useful for future users.
Additional informations on the use of the iCATS tool, can be found in the git-hub repository here: \url{https://github.com/dimihayl/DLM/tree/devImCats}.\\
First of all, important to mention is the way that iCATS handles internally the complex potential. As shown in Ref.~\cite{Kamiya:DD}, when obtaining the solution of the \SE equation with a complex potential in a femtoscopy setup, the wave function must satisfy the asymptotic boundary condition where the outgoing wave is normalized to unity. To achieve this requirement, the strong complex potential $V_l(r) = V_{R,l} (r)+i  V_{I,l} (r)$ written in terms of its real and imaginary component and for a specific partial wave $l$ enters in the radial equation with its hermite conjugate $V_l(r) ^\dagger$.\\
The iCATS iterative numerical procedure to solve the coupled system in Eq.~\ref{eq:System} is based on a \textit{finite difference method} (FDM) and a three grid points  in $r$ ($[i-1,i,i+1]$).
To simplify the notation, we will label in the appendix:
\begin{align}
    & u_R(r) = f(r), \,\, u_I(r)=g(r)\nonumber \\
    & \mathcal{F}_1= 2m V_R +\frac{l(l+1)}{r^2} -k^2 \\
    & \mathcal{F}_2= 2m V_I 
\end{align} 
Within this notation, the system of equations to be solved reads:
\begin{align}\label{eq:fgSystem}
    \begin{cases}
        f^{''} = \mathcal{F}_1 f +\mathcal{F}_2 g \\
        g^{''} = \mathcal{F}_1 g - \mathcal{F}_2 f. \\
    \end{cases}
\end{align}
The $\mathcal{F}_{1,2}$ quantities are evaluated at each grid point. 
The basic idea is to determine the values of $f$ and $g$ at each subsequent point, knowing their values in the previous two r-grid points. An initial value for both $f$ and $g$ at points $i=0$ and $i=1$ is assumed (given by the free spherical Bessel), then the evaluation proceeds to estimate the two functions at $i+1$ until convergence is reached.\\
In order to solve Eqns.~\ref{eq:fgSystem} we adopt, similarly to CATS, a Euler discretization with an adaptive grid. The discretized version reads
\begin{align}\label{eq:fgSystemEulerFinal}
    & f_{i+1} = f_i \left( 1+\dfrac{\Delta_i}{\Delta_{i-1}}\right) - f_{i-1}\dfrac{\Delta_i}{\Delta_{i-1}} +\Delta_i^2 (f_i \cdot \mathcal{F}_{1,i} +g_i \cdot \mathcal{F}_{2,i} )\\
    & g_{i+1} = g_i \left( 1+\dfrac{\Delta_i}{\Delta_{i-1}}\right) - g_{i-1}\dfrac{\Delta_i}{\Delta_{i-1}} +\Delta_i^2 (g_i \cdot \mathcal{F}_{1,i} -f_i \cdot \mathcal{F}_{2,i} ).
\end{align}
where $\Delta_i = r_{i+1}-r_i$. Additionally, the current iCATS implementation for the asymptotic condition allows to easily extend the calculations to include possible two- and three-coupled channels interactions.\\

\section{Additional plots}\label{Appendix B.}
\subsection{\LAL and \pAL correlation fits in \pp collisions}
\begin{figure*}[t!]
\begin{center}
\includegraphics[width=0.95\columnwidth]{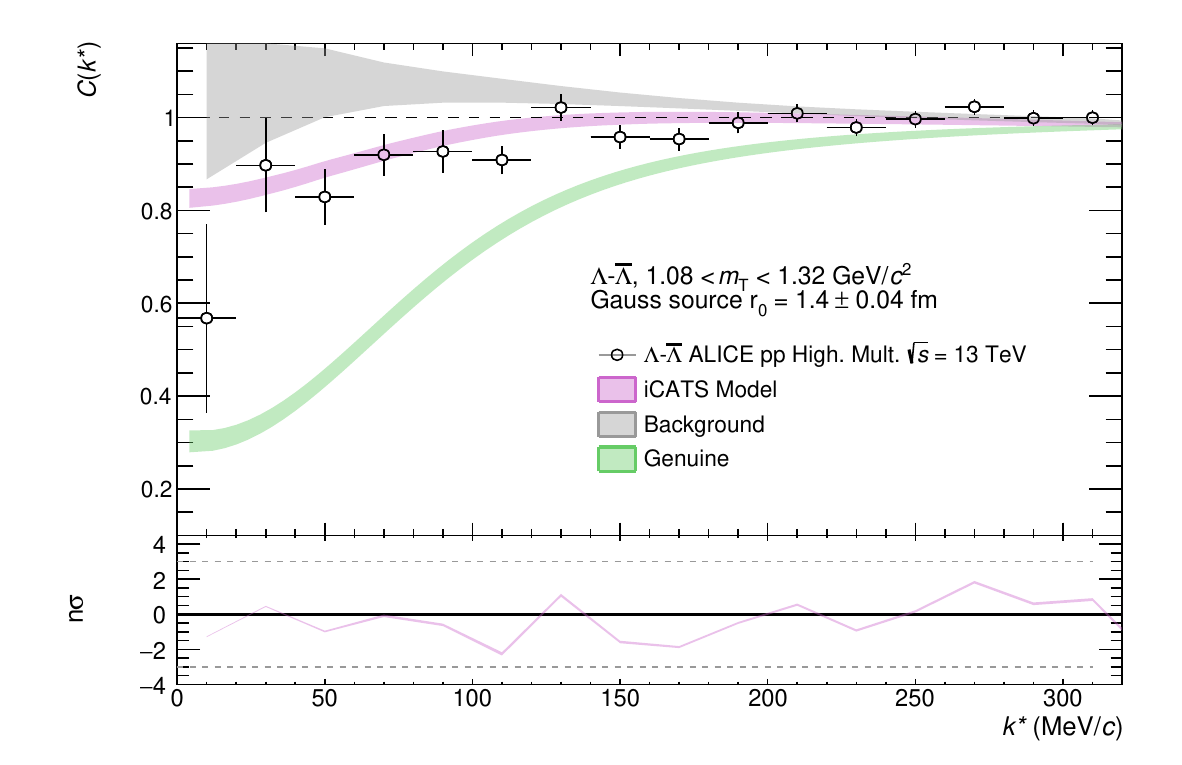}
\includegraphics[width=0.95\columnwidth]{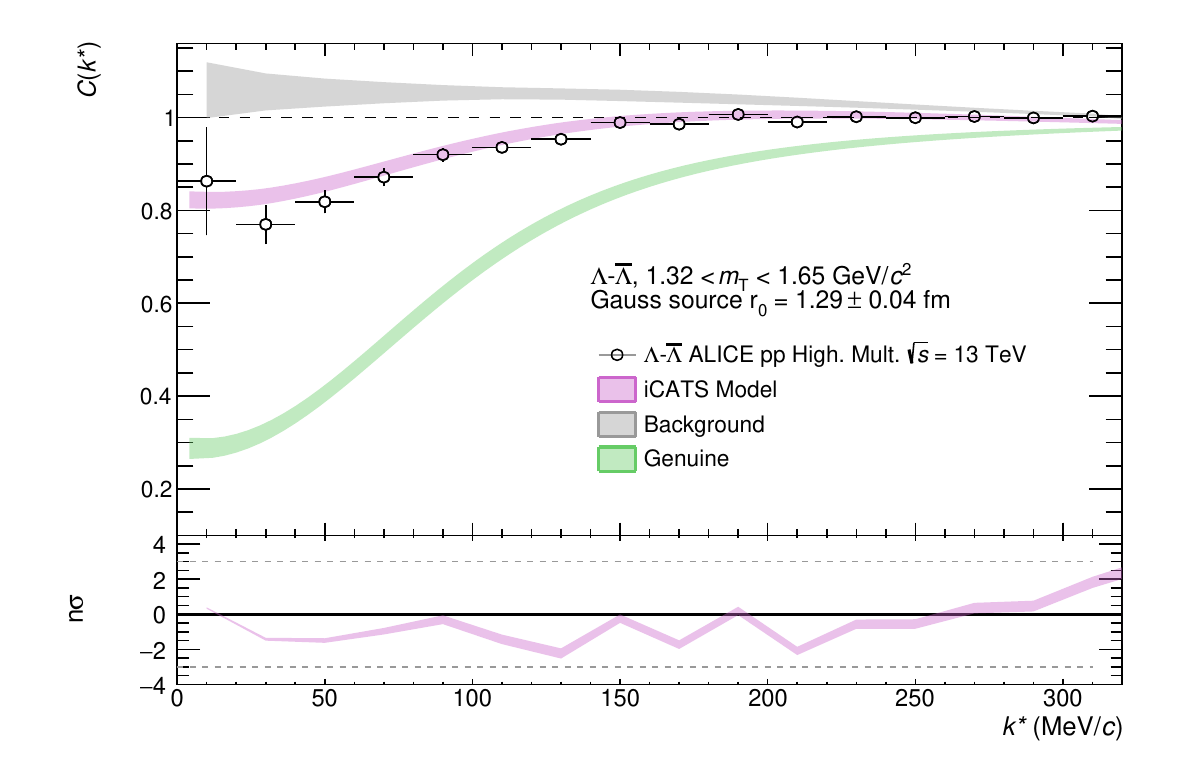}
\caption{(Color online). Results obtained with iCATS (pink band) for the \LAL measured correlation functions. The publicly available data (black empty markers) and background (grey band) correspond to the ones in Fig.A.2 of Ref.~\cite{ALICE:BBarpp}.}
\label{figApp:LALCF_pp}
\end{center}
\end{figure*}

\begin{figure*}[h!]
\begin{center}
\includegraphics[width=0.91\columnwidth]{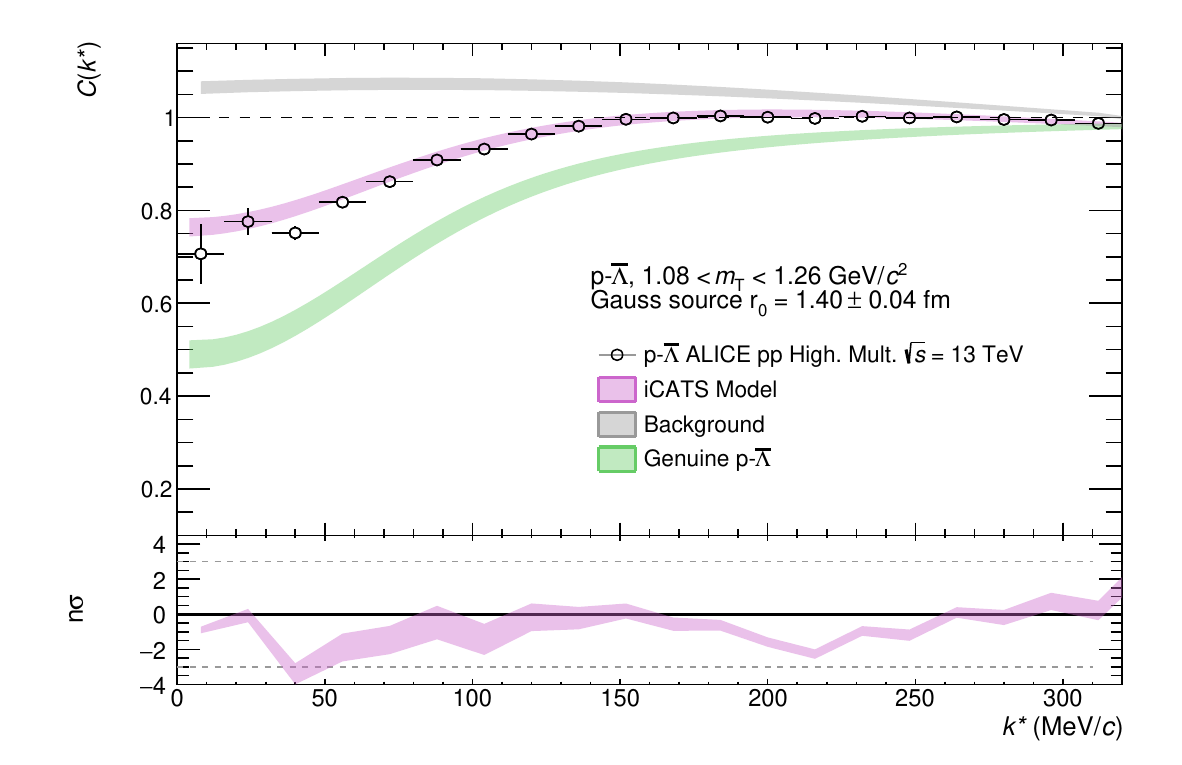}
\includegraphics[width=0.91\columnwidth]{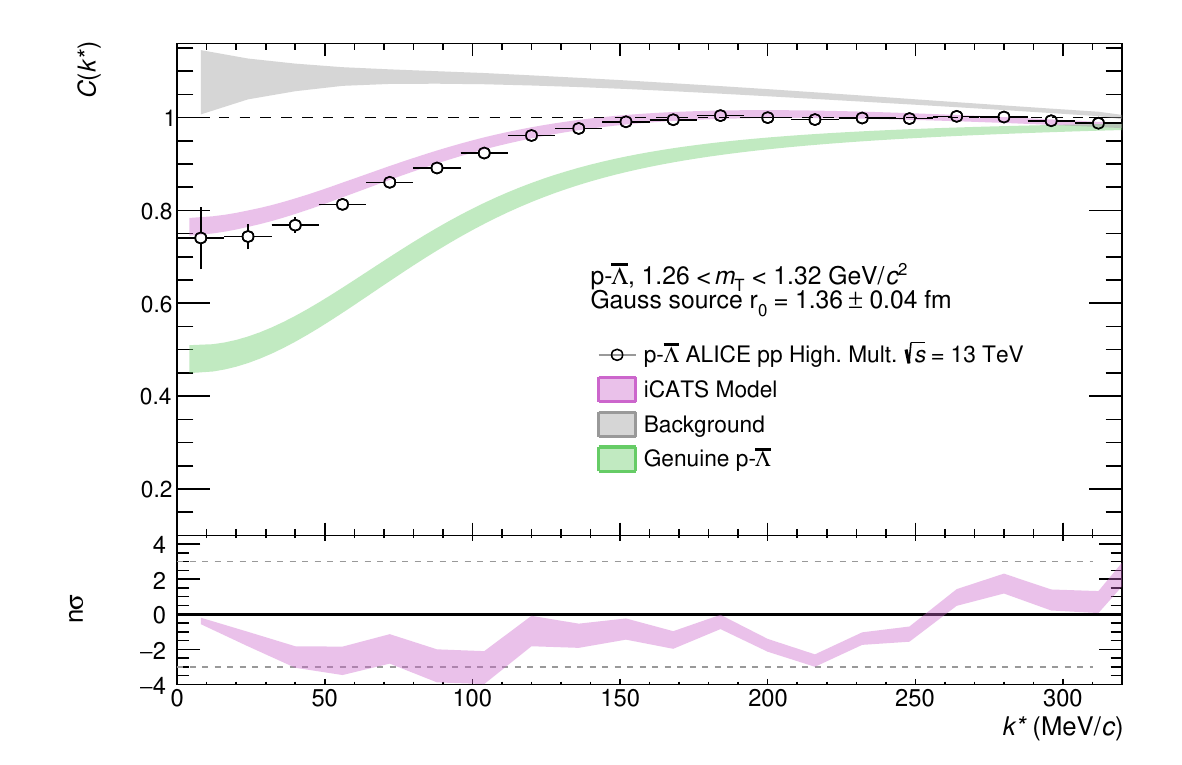}\\
\includegraphics[width=0.91\columnwidth]{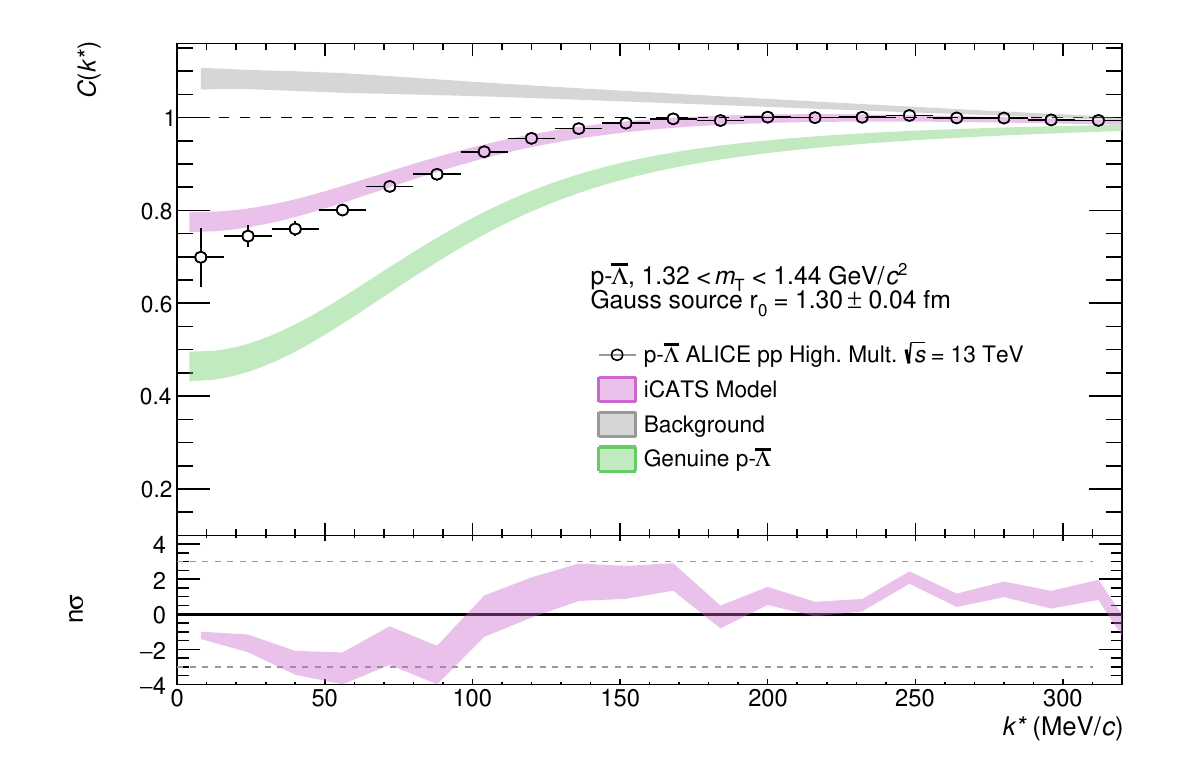}
\includegraphics[width=0.91\columnwidth]{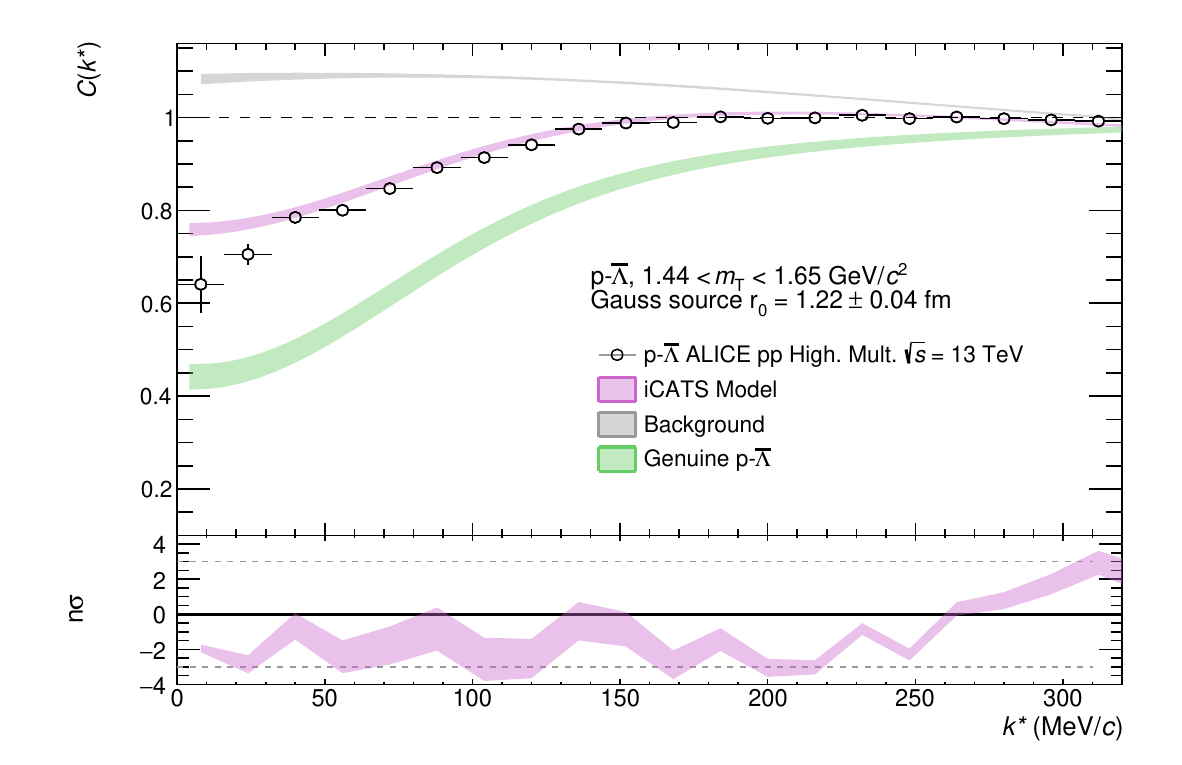}\\
\includegraphics[width=0.91\columnwidth]{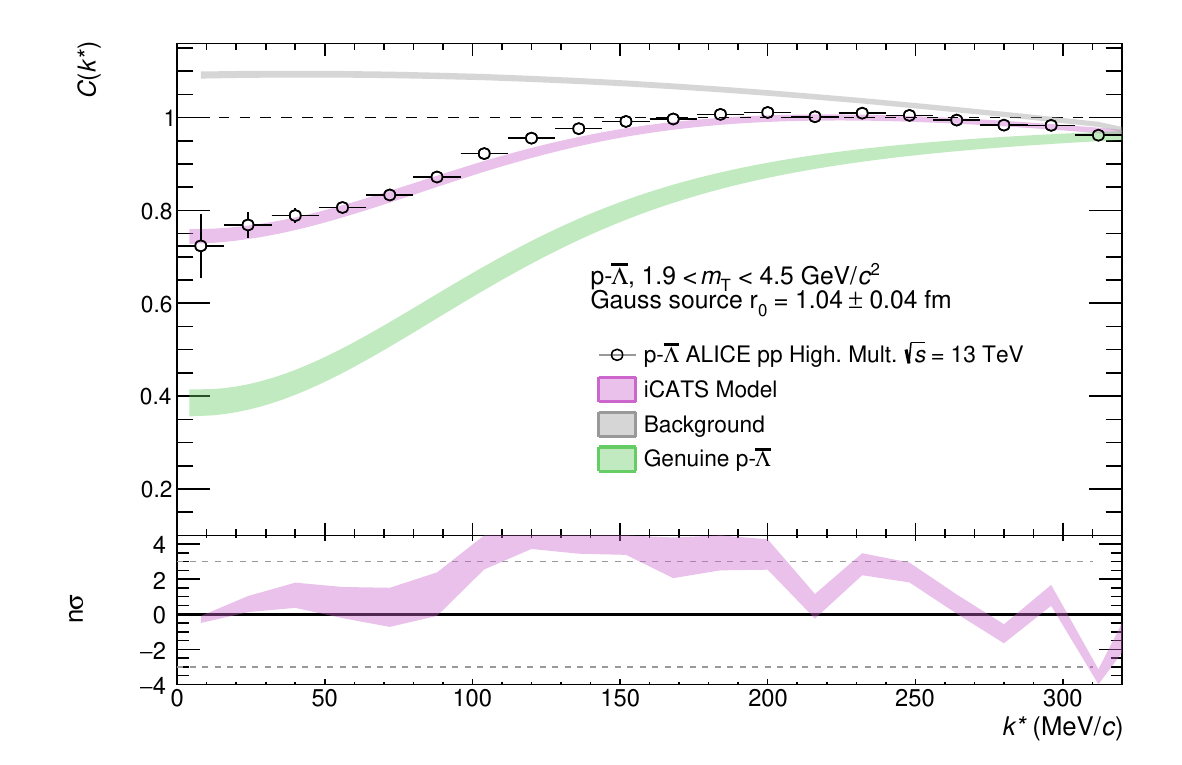}
\caption{(Color online). Results obtained with iCATS (pink band) for the \pAL measured correlation functions. The publicly available data (black empty markers) and background (grey band) correspond to the ones in Fig.A.1 of Ref.~\cite{ALICE:BBarpp}. }
\label{figApp:pALCF_pp}
\end{center}
\end{figure*}

In Fig.~\ref{figApp:LALCF_pp} and Fig.~\ref{figApp:pALCF_pp} we present the remaining \mt fit results on the \LAL and \pAL pairs.
Values of the effective emitting sources for these \mt bins have been obtained via private communication with the ALICE analysers and reported here:
\begin{table}[h!]
\caption{Effective Gaussian radii $r_{0}$ used in the femtoscopic fit in this work and in~\cite{ALICE:BBarpp}.}
\begin{center}
\begin{tabular}{ c|c|c } 
\hline
Pair & \mt range & $r_{0}$ (fm) \\
  \hline
  \multirow{3}{4em}{\pAL} 
   & $1.08-1.26$  & $1.4\pm 0.04$\\ 
  & $1.26-1.32$ & $1.36\pm 0.04$ \\ 
  & $1.32-1.44$ &  $1.3\pm 0.04$\\
  & $1.44-1.65$ &  $1.22\pm 0.04$\\ 
  &  $1.9-4.5$ &  $1.04\pm 0.04$\\ 
  \hline
  \multirow{3}{4em}{\LAL} 
  & $1.08-1.32$ & $1.4\pm 0.04$\\ 
  & $1.32-1.65$ & $1.29\pm 0.04$ \\ 
  \hline
\end{tabular}
\end{center}
\label{tab:effgaussiansource}
\end{table}



Results on the measured \PbPb correlation data obtained in~\cite{ALICE:BBarPbPb}. The reported values of the radii have been obtained via private communication with the ALICE analysers and reported on the figures.
\begin{figure*}[h!]
\begin{center}
\includegraphics[width=\columnwidth]{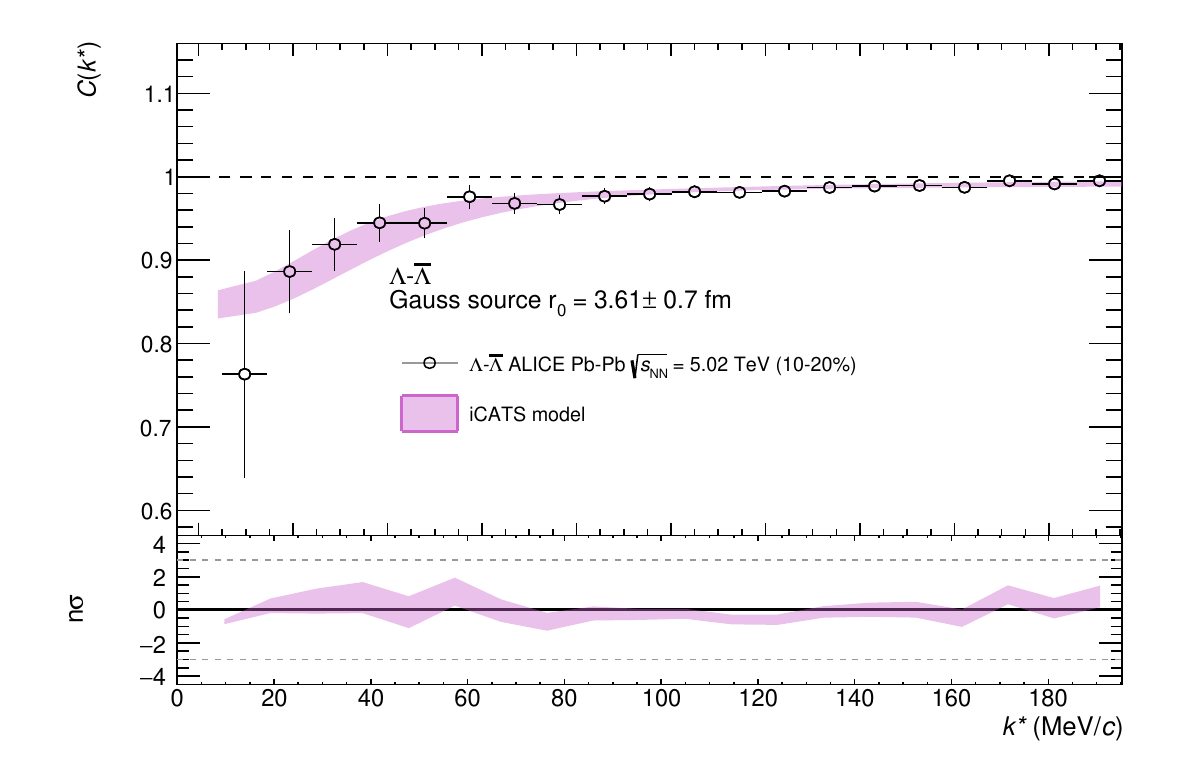}
\includegraphics[width=\columnwidth]{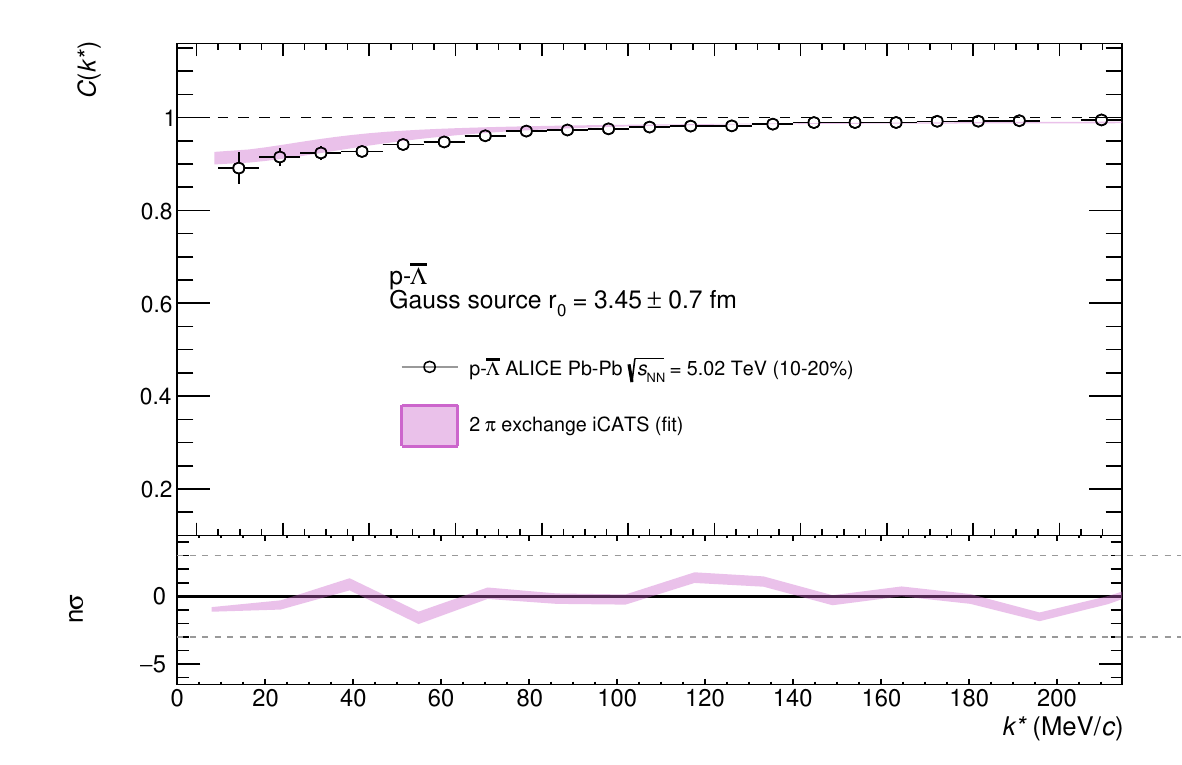}
\caption{(Color online). Results obtained with iCATS for the \pAL and \LAL measured correlation functions in \PbPb at \fivenn.  The lower panels deliver the deviation, expressed in terms of standard deviations $\sigma$, of the fit results to the data}
\label{figApp:CFPbPb}
\end{center}
\end{figure*}

\begin{figure*}[h!]
\begin{center}
\includegraphics[width=\columnwidth]{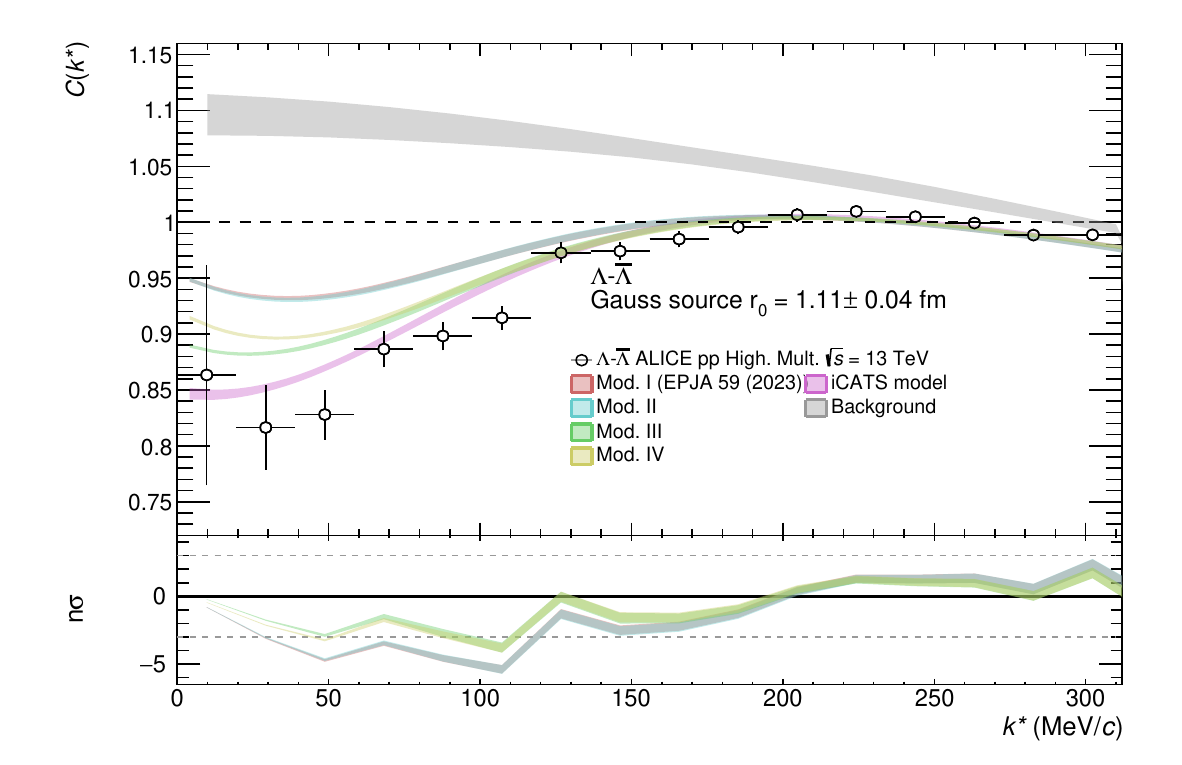}
\includegraphics[width=\columnwidth]{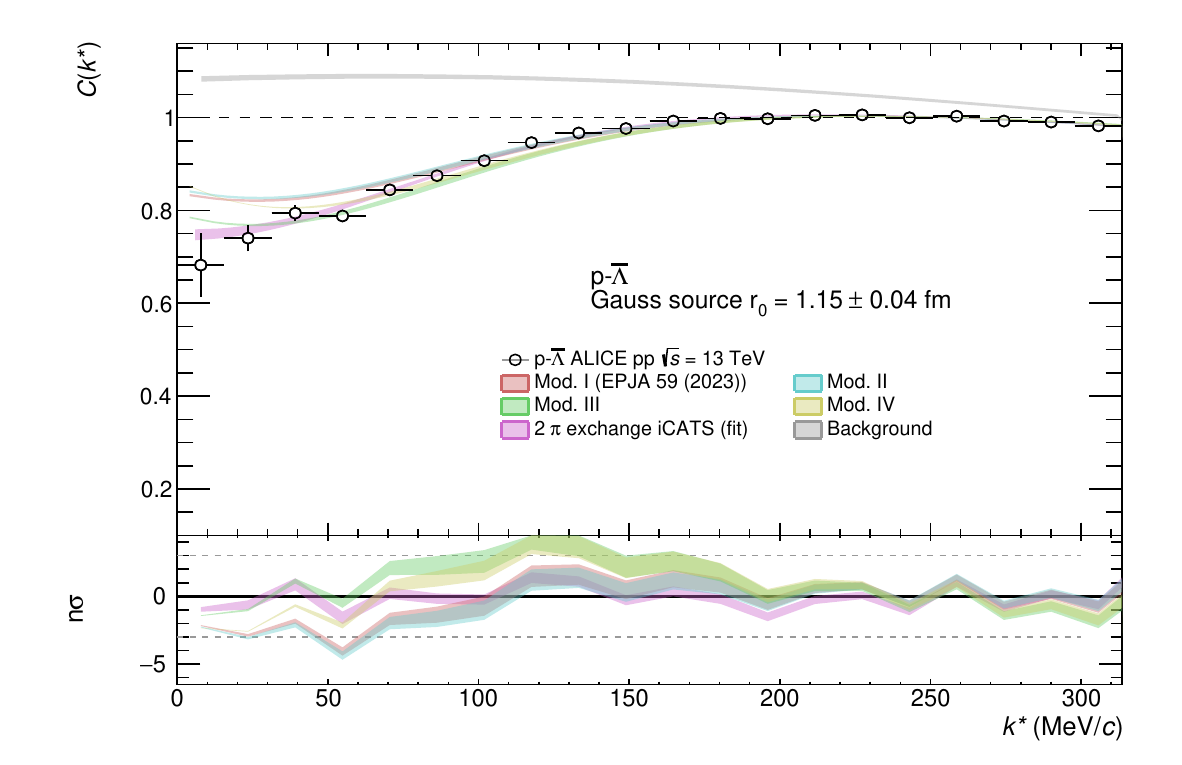}
\caption{(Color online). Results obtained with iCATS for the \pAL and \LAL measured correlation functions assuming the meson-exchange models (I,II,III,IV) in Ref.~\cite{Haidenbauer:LAL}.  The lower panels deliver the deviation, expressed in terms of standard deviations $\sigma$, of the fit results to the data}
\label{figApp:JohannComparisonModels}
\end{center}
\end{figure*}

In Fig.~\ref{figApp:JohannComparisonModels} we report the results, on the \LAL (left) and on the \pAL pairs (right) using optical potentials in iCATS tuned to reproduce the scattering lengths reported in Tab.2 of Ref~\cite{Haidenbauer:LAL}. 

\subsection{$\pAp \rightarrow \LAL$ reaction cross-section}
\begin{figure}[h!]
\begin{center}
\includegraphics[width=\columnwidth]{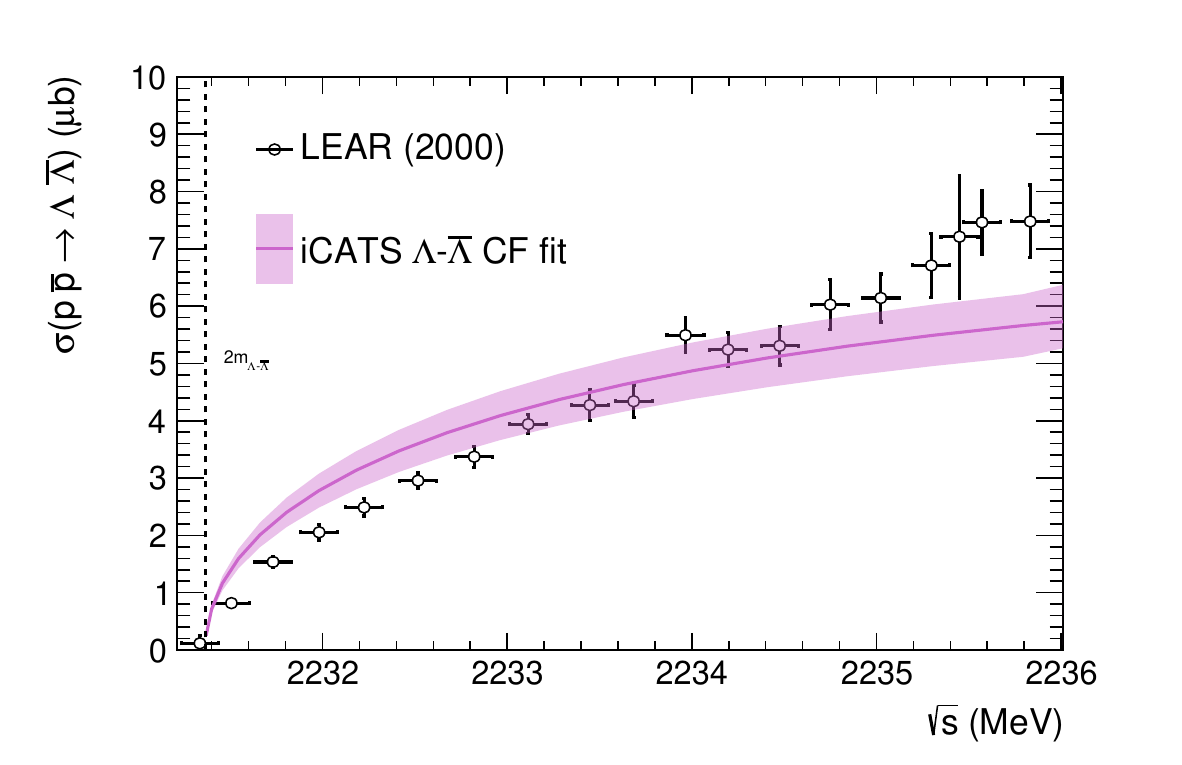}
\caption{(Color online). Measured $\pAp \rightarrow \LAL$ cross-section from the PS185 experiment at LEAR~\cite{Barnes:2000be}. Results obtained with the \LAL iCATS interaction are shown in the pink band.}
\label{figApp:LEAR}
\end{center}
\end{figure}


\clearpage  

\bibliographystyle{unsrtnat}
\bibliography{bibliography.bib}
\end{document}